\documentclass{optica-article}
\journal{opticajournal}

\articletype{Research Article}
\usepackage{lineno}
\usepackage{mathtools}
\usepackage{xcolor}

\begin{document}

\title{Scalar computational primitives with perturbative phase interferometry}

\author{Christopher R. Schwarze,\authormark{1, *}, Anthony D. Manni,\authormark{1} David S. Simon,\authormark{1,2} and Alexander V. Sergienko\authormark{1,3}}
\address{\authormark{1}Department of Electrical and Computer Engineering \& Photonics Center, Boston University, 8 Saint Mary’s St., Boston, Massachusetts 02215, USA\\
\authormark{2}Department of Physics and Astronomy, Stonehill College, 320 Washington Street, Easton, Massachusetts 02357, USA\\
\authormark{3}Department of Physics, Boston University, 590 Commonwealth Avenue, Boston, Massachusetts 02215, USA}
\email{\authormark{*}crs2@bu.edu}

\begin{abstract}
We describe how weak phase modulations applied to classical coherent light in specially modified linear interferometers can be used to perform primitive computational tasks. Instead of encoding operations within a fixed unitary state, the operations are enacted by moving from one state to another. This harnesses the particular phase parametrization of an interferometer, allowing entirely linear optics to produce nonlinear operations such as division and powers. This is due to the nonlinear structure of the underlying phase parametrizations. The realized operations are approximate but can be made more accurate by decreasing the size of the input perturbations. For each operation, the inputs and outputs are changes in phase relative to a fixed bias point. The output phase is ultimately read out as a change in optical power.
\end{abstract}

\section{Introduction}
Alongside modern efforts in traditional quantum computing with photonic qubits, much attention is being directed toward optical computing with classical light sources. Broadly speaking, this analog alternative is thought to be promising due to its potential to increase the speed, bandwidth, and energy efficiency of specific computing tasks relative to digital electronic computers \cite{McMahon2023}. A review of modern approaches is given by Ref. \cite{StroevReview}.

Certainly light possesses some concrete characteristics that appear amenable to building useful computing systems, such as the interference of amplitudes, the many orthogonal information-carrying modes which can be accessed, and the ability to carry information great distances in minimum time. Nonetheless a number of challenges need to be addressed, such as the programming interface, limitations on realizable operations, accumulation of errors, and information bottlenecks that could drastically hinder the overall compute throughput. 

A particular class of computations of widespread interest is matrix-vector multiplication \cite{Athale:82, 10.1063/5.0235712, 10.1117/12.2028585, Gruber:00, Spall:20, Tang:25} in part due to its heavy and ever-increasing role in contemporary approaches to machine learning \cite{Shen2017, Wang2022, Filipovich:22, Hua2025, doi:10.1126/sciadv.ads4224, Ahmed2025}. Certain optical techniques for matrix-vector multiplication include spatial-light modulation \cite{StroevReview, doi:10.1126/sciadv.adg7904, PhysRevApplied.15.054034}, photonic meshes \cite{Bogaerts2020, Nakajima2021}, reprogrammable multimode waveguides \cite{onodera2024scalingonchipphotonicneural}, and photoelectric multiplication \cite{PhysRevX.9.021032}. In many of these approaches, the $n \times n$ matrix transformation $U_0$ is obtained by tuning some $m$ internal phases $\Phi = (\phi_1, \phi_2, \dots, \phi_m)$ of an interferometer to realize the scattering action $U(\Phi)$. This automatically provides a basis for programmability, but introduces two related complications. The first is that the programming can be nontrivial due to the indirect encoding between the elements of $\Phi$ and the elements of $U(\Phi)$; algorithms are often required to load a desired matrix $U_0$ using an iterative procedure that uses feedback in the form of measurements of $U(\Phi)$. The procedure ends when $U(\Phi)$ is sufficiently close to $U$. Once the phases $\Phi$ are fixed, the computation itself is fast. But when the target $U_0$ needs to be changed, the system needs to be placed at a new bias point in $\Phi$, and the associated reconfiguration time can be costly.

$U_0 \approx U(\Phi)$ is a single point in the parameter space of $U$ spanned by $\Phi$. Errors in the construction of $U$ and/or selection of a desired $\Phi$ lead to deviations from the obtained matrix $U(\Phi)$ and the desired matrix $U_0$, which in effect produces a different computation than what was targeted. The indirect encoding is also responsible for an undesirable accumulation of errors, as a single $\phi_j$ can affect many elements of $U_0$. Another aspect of positioning the system at just one point in the parameter space spanned by $U(\Phi)$ is that it discards the device's particular parametrization of the phases $\Phi$ in that space, which is due to the graphical structure of the interferometer. For a general network of scatterers forming another aggregate, phase-tunable scatterer, the parametrization is highly dependent on both the properties of the constituent scatterers and their interconnect topology. 

For a fixed wavelength and polarization, the local geometry of some neighborhood of scattering states in a given interferometer's parameter space may be understood in terms of a local expansion around some point $\Phi_0$. The expansion coefficients dictate physical quantities such as phase responsivity and the frequency dispersion coefficients. Unlike the individual values of the state amplitudes at a given state (of a fixed frequency and polarization), these quantities are tied to the particular parametrization used. Different wavelengths and/or polarizations will have their own associated parameter spaces which in general can be quite different in character; broadband and non-polarizing devices are designed to drive these differing modal parameter spaces toward coalescence. 

Certain phase parametrizations have desirable properties which can only be harnessed as a resource for computation if multiple states in a neighborhood of the state space are involved in the same computational procedure. A more elaborate description of this will be provided in the next section. In general, the considered phase modulation technique could be electrical, thermal, optical (via self-phase modulation or cross phase modulation) etc. Crucially, instead of designing computations enacted at a particular point of parameter space, \textit{we propose a scheme which enacts computation by a change in state from one point to another in some compact subset of the parameter space.} Then, allowing the magnitude of the change to be a continuously varying parameter, the various subsets of the parameter space considered become the grounds of different operations. 

\begin{figure}
    \centering
    \includegraphics[width=0.75\linewidth]{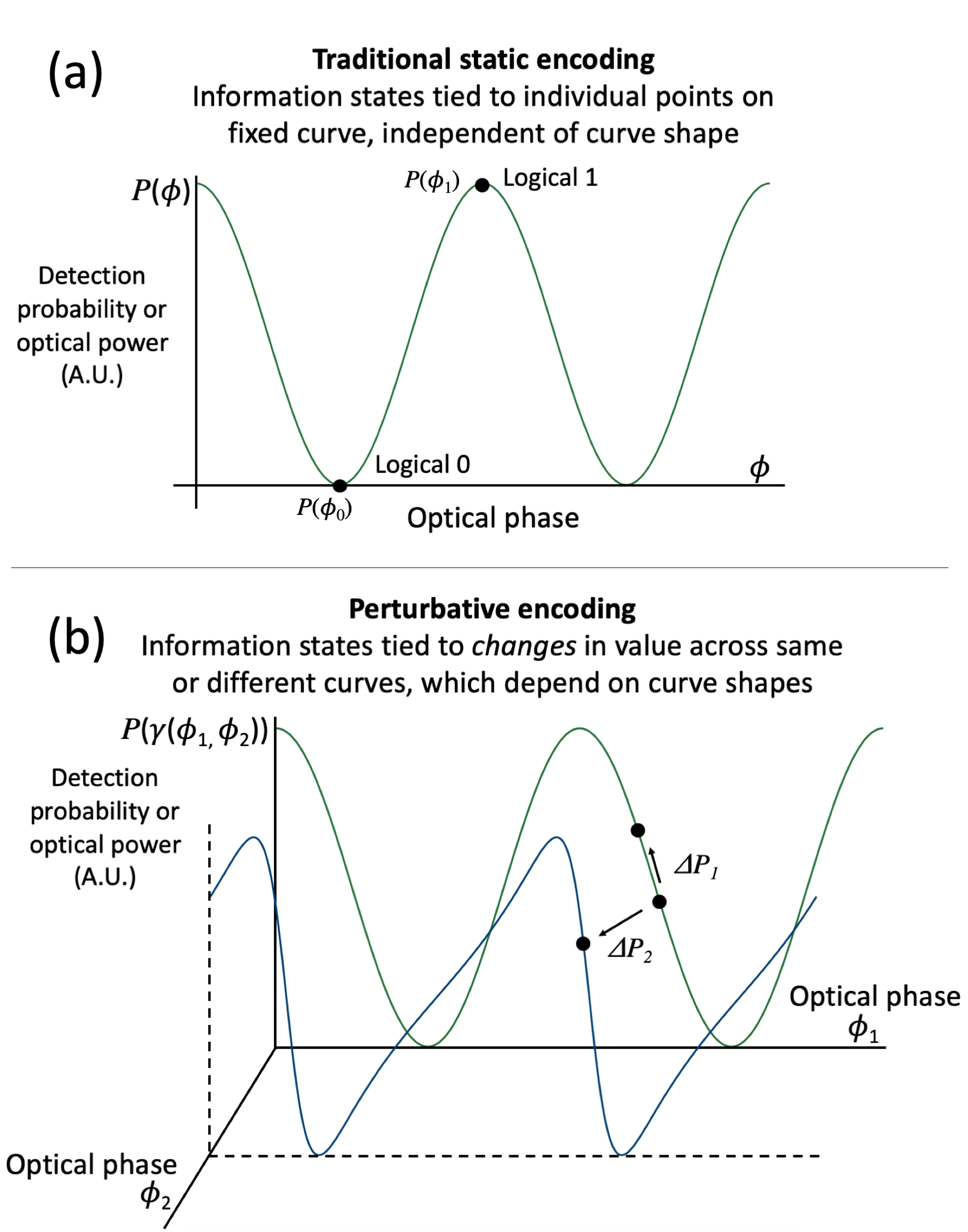}
    \caption{A comparison of interferometric encoding schemes. The traditional approach (top) assigns information values to individual points in the parameter space. This ties information to the optical state (i.e. photon number distribution among various spatial modes) but discards the geometric information in the parameter space. The proposed approach (bottom) ties the information to a \textit{change} in the optical state induced by a change in one or more phases. In coupled-phase interferometers, changing one phase can deform the curve which is defined by spanning another phase. In this encoding approach, the nature of the deformation can now be used to enact certain information processing operations. That is, $\Delta P$ represents a computed output for a given set of inputs $(\Delta \phi_1, \Delta \phi_2)$. In particular, $\Delta P$ is intended to directly read out the change in phase $\Delta \gamma(\Delta \phi_1, \Delta \phi_2)$, so all computations are entirely encoded in the optical phase. Examples derived in the text include addition and subtraction (Sec. 3.1), multiplication (Sec. 3.2), nonlinear operations such as squaring or scalar inversion (Sec. 3.3), and cascaded application of these (Sec. 3.4).}
    \label{fig:concept}
\end{figure}

The proposed optical computing scheme is conceptually compared to the traditional, fixed-state encoding in Fig. \ref{fig:concept}. With our proposed encoding, we will illustrate how the four arithmetic primitives can be obtained from the same basic interferometric configuration. The scheme uses entirely linear optics yet is able to produce nonlinear operations such as division and powers due to the nonlinear nature of the parametrizations considered. Although this approach is presently used to produce a variety of scalar operations, it can be extended to matrix and vector operations, which will be left for a future work. 

In the next section, more details and examples are provided on the notion of a scattering state space parametrization. A brief review of the relevant background on the linear-optical scattering matrix formalism can be found in Section 2 of Ref. \cite{SchwarzeJB}. In Section 3, the perturbative phase optical computing scheme is described for individual and cascaded scalar operations. Throughout this article, we will adopt a particular phase parametrization to illustrate the concept, which is derived in the next section. This parametrization is chosen since it is prevalent, versatile, and straightforward to realize. Still, other undiscovered parametrizations may exist which are more accurate over a larger input range. In the final section, we discuss the results, consider possible extensions and adaptions of this approach, draw connections to related work, and summarize final conclusions.

\section{Scattering state space parametrizations}

Let the set of linear field scattering transformations spanned by a phase-tunable device $U$ be $\mathcal{U} = \{U(\phi_1, \phi_2, \dots, \phi_m)\ |\ (\phi_1, \dots, \phi_m) \in \mathbb{C}^m \}$. For lossless systems, $\mathcal{U}$ is a subset of the unitary group $U(n)$ and the phase parameters lie in $\mathbb{R}^m$. In this case, the phases may be restricted to lie in the set $[-\pi, \pi]^m$, $[0, 2\pi]^m$, or any shifted version thereof without loss of generality. 

The set $\mathcal{U}$ can be referred to as $U$'s ``state space'' because it describes the different optical states that $U$ generates when its phase parameters are swept. A state $|\psi_{\text{in}}\rangle$ incident to $U$ will be mapped to $|\psi_{\text{out}}\rangle = U(\phi_1, \dots, \phi_m)|\psi_{\text{in}}\rangle$. The state space can be visualized by treating each column of $U$ as a vector in $\mathbb{C}^n$ and tracing the paths these vectors sweep as the parameters $\Phi$ are varied. If the input phases to a device $U$ are re-parametrized by a continuous map $\gamma$ sending $(\phi_1, \dots, \phi_m)$ to $\gamma(\phi_1, \dots, \phi_m)$ the space $\mathcal{U}$ itself does not change, but the manner in which these paths are traced will change. This implies that the manner in which the device responds to physical modifications of the phase parameters will change as well. However, if the system is confined to a single fixed state, the underlying parametrization is irrelevant. 

\begin{figure}[htb!]
    \centering
    \includegraphics[width=0.5\linewidth]{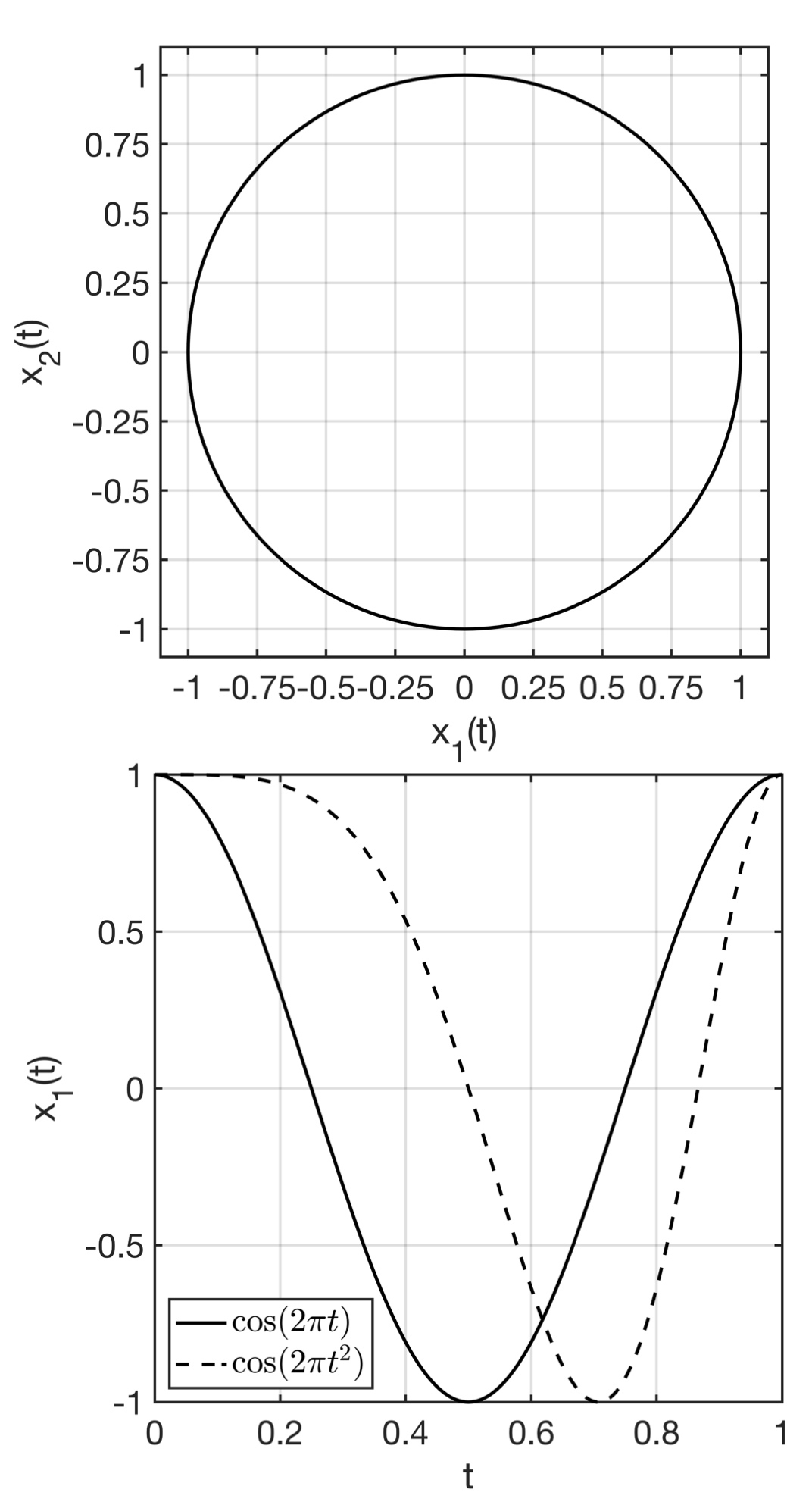}
    \caption{Geometric objects such as the circle plotted on the top may be parametrized in different ways. The horizontal component of two example parametrizations of this circle, $\mathbf{f}_1(t) = (\cos(2\pi t), \sin(2\pi t))$ and $\mathbf{f}_2(t) = (\cos(2\pi t^2), \sin(2\pi t^2))$ for $0\leq t \leq 1$ are shown on the right \cite{guggenheimer}. In the same sense, optical interferometers can parametrize a given scattering state space in different ways. This is exemplified in the Michelson and Grover-Michelson interferometers compared in Fig. \ref{fig:migmi}.}
    \label{fig:reparam-example}
\end{figure}

The function $\gamma$ is the phase outputted by an interferometer, which depends on the structure of the interferometer itself. The phases $\phi_1, \dots, \phi_m$ are the user-controllable input phases. When $U(\phi_1, \dots, \phi_m)$ is replaced by $U(\gamma(\phi_1, \dots, \phi_m))$ any tunable scattering system concerned only with individual points will be unaltered, but a perspective concerned with the neighborhoods around various points will witness a change in the structure of these neighborhoods. As a textbook example, the same goes for any parametrized plane curve $\mathbf{f}(t) = (x_1(t), x_2(t))$ when the parametrization of $t$ changes \cite{guggenheimer}. As indicated in Fig. \ref{fig:reparam-example}, the plane curves $\mathbf{f}_1(t) = (\cos(2\pi t), \sin(2\pi t))$ and $\mathbf{f}_2(t) = (\cos(2\pi t^2), \sin(2\pi t^2))$ for $0\leq t \leq 1$ are different parametrizations of the same geometric object, the unit circle. $\mathbf{f}_2$ can be recognized as $\mathbf{f}_1$ subject to the reparametrization $\gamma(t) = t^2$. 

\begin{figure}[htb!]
    \centering
    \includegraphics[width=0.6\linewidth]{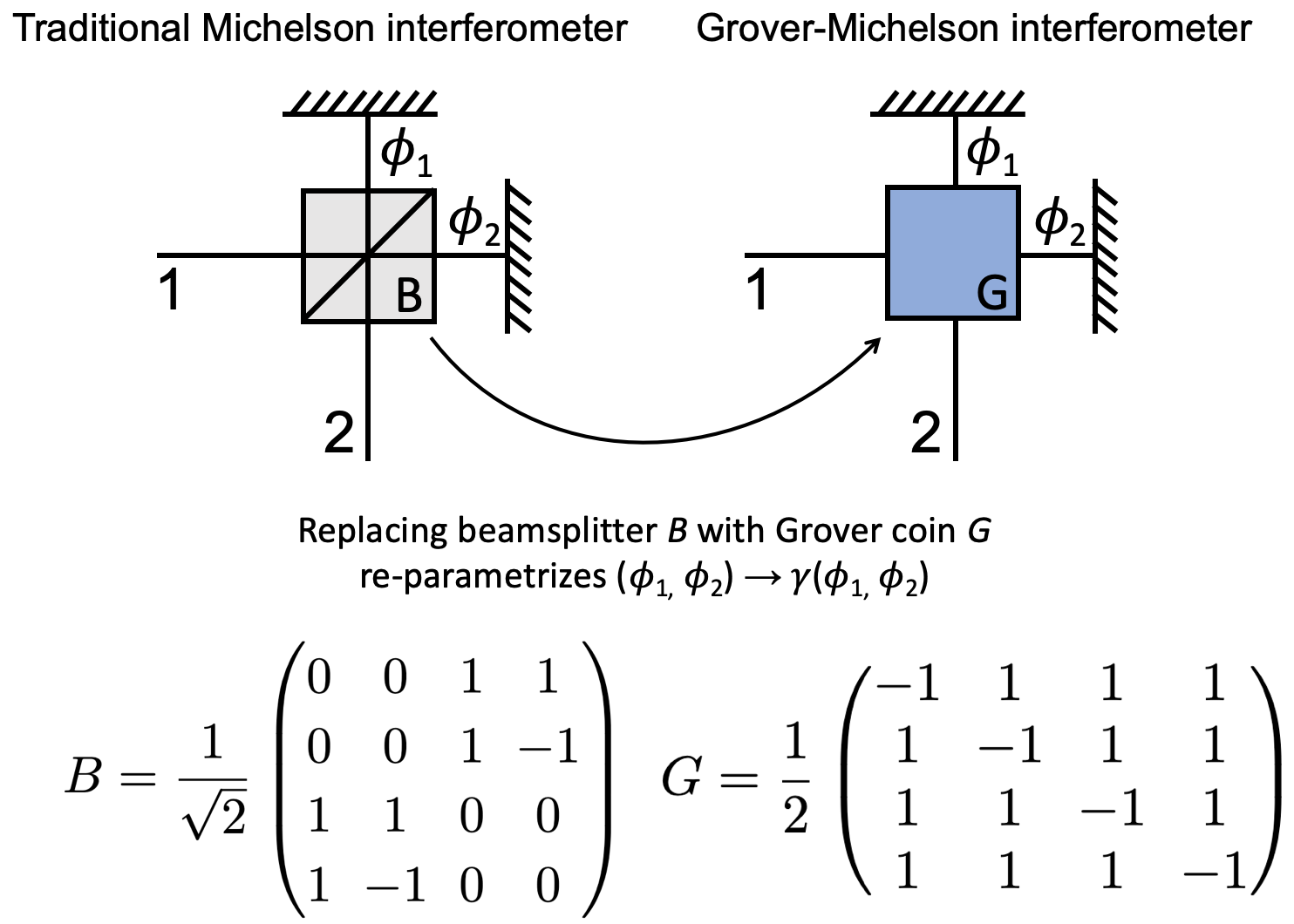}
    \caption{The traditional Michelson interferometer (left) is formed with a beam splitter $B$ and two end-mirrors. When light scatters into $B$, all of the light passes from the input ports 1 and 2 to the mirror-arms; there is no back-scatter or light coupled between ports 1 and 2. Thus the light enters through the mirror-arms, acquires an optical phase of $\phi_j$ in arm $j$, and exits the device in one pass. However, replacing the beam-splitter with a four-port Grover coin $G$ now allows light to scatter from one port to another with an equal probability of $25\%$. The resulting device, the Grover-Michelson interferometer (right), is now a multi-pass system of coupled resonator mirror-arms, and is governed by an effective phase $\gamma(\phi_1, \phi_2)$ \cite{PhysRevA.107.052615}. This mapping $\gamma$ is a nonlinear reparametrization of the physical phases $(\phi_1, \phi_2)$. It is explicitly given by $\gamma(\phi_1, \phi_2) = \arctan((\sin(\phi_1 + \phi_2) - \sin\phi_1 - \sin\phi_2 )/(\cos(\phi_1 + \phi_2) - \cos\phi_1 - \cos\phi_2 + (1 + \cos^2(\phi_1 - \phi_2))/2))$. Additional plots of these different parametrizations can be found in Fig. 5 of Ref. \cite{PhysRevA.107.052615} and experimentally in Fig. 6 of Ref. \cite{Schwarze:24}.}
    \label{fig:migmi}
\end{figure}

In the same way that different curve parametrizations can represent the same geometric object, different interferometers may provide different parametrizations of the same scattering state space. After all, each element of a tunable $U$ also traces a curve in the complex plane as its parameters are varied. An example of this distinction in the context of interferometry is given by the Michelson and Grover-Michelson interferometers. These devices are depicted schematically in Fig. \ref{fig:migmi}. The Grover-Michelson is obtained when the beam-splitter of a traditional Michelson interferometer is replaced by a certain generalization which produces four beams of equal intensity instead of the regular two \cite{PhysRevA.107.052615, Schwarze:24}. This amounts to employing the Grover scattering matrix
\begin{equation}
    G = \frac12
    \begin{pmatrix}
        -1 & 1 & 1 & 1\\
        1 & -1 & 1 & 1\\
        1 & 1 & -1 & 1\\
        1 & 1 & 1 & -1
    \end{pmatrix},
\end{equation}
in place of the regular beam-splitter matrix, 
\begin{equation}
    B = \frac{1}{\sqrt{2}}
    \begin{pmatrix}
        0 & 0 & 1 & 1\\
        0 & 0 & 1 & -1\\
        1 & 1 & 0 & 0\\
        1 & -1 & 0 & 0
    \end{pmatrix}.
\end{equation}
The substituted scatterer $G$ has $1/4$ probability for light to back-scatter and $1/4$ probability for transmitting to each of the other ports. This entails the formation of coupled resonators in the arms of the interferometer, leading to a nonlinear response in the round-trip phases $\phi_1$ and $\phi_2$ which is not present in the traditional Michelson interferometer. In the Michelson interferometer, the optical power output is sinusoidal in $\phi_1$ for any $\phi_2$, \textit{whereas in the Grover-Michelson system, each value of $\phi_2$ produces an entirely new curve in $\phi_1$.}

The two devices both cover the space of two-port, reciprocal unitary scatterers. This equates to spanning the line $R + T = 1$ where $R$ is the reflection probability and $T$ is the transmission probability. This means both are able to act as a two-port beam-splitter of any desired splitting ratio, or more generally, produce any $U(2)$ scattering device up to external phase shifts. However, the different phase parametrizations possessed by each device generate different curves $R$ and/or $T$ vs. $\phi_1$ for various fixed $\phi_2$. In turn, these different curves exhibit regions of varying phase sensitivity and produce fringes of different shapes. In particular, $\phi_2$ will translate the sinusoid produced by a Michelson interferometer, but will continuously deform the curves of a Grover-Michelson, producing curves with regions of very flat and very sharp slopes, as depicted in Fig. 5 of Ref. \cite{PhysRevA.107.052615}.

Next we will consider two-port scatterers which are fully transmitting. These devices can only enact a phase transformation on the light passing through it: $|\psi_{\text{out}} \rangle = e^{i\gamma} |\psi_{\text{in}}\rangle$. This leads to a generalized notion of a phase shift, where the phase $\gamma$ acquired by the incident state is a reparametrization of the physical phase $\phi$ applied within the device. Certain $\gamma = \gamma(\phi)$ can amplify small changes in the inputted physical phase into effectively larger ones without changing the intensity or direction of the light passing through \cite{PhysRevA.109.053508}. Thus, this class of parametrizations can behave as a linear amplifier of optical phase. As will be exemplified shortly, it is straightforward to construct systems where the parametrization itself can be continuously tuned by an auxiliary phase, which will be denoted $\delta$. For each value of $\delta$, a new reparametrization is given by the map $\phi \rightarrow \gamma(\phi, \delta)$. In that context $\gamma$ is promoted from function to homotopy.

\begin{figure}
    \centering
    \includegraphics[width=0.6\linewidth]{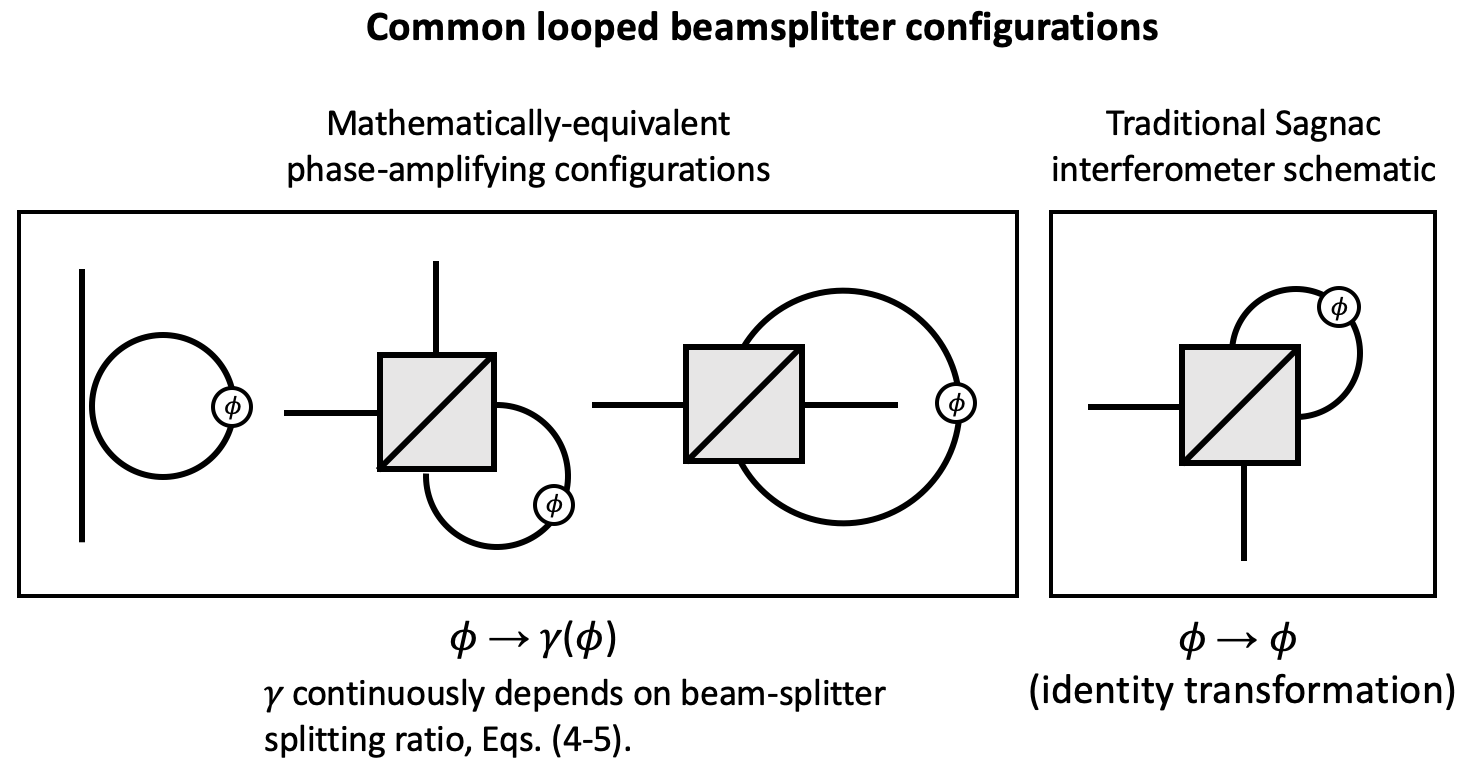}
    \caption{Different forms of a looped beam-divider, including the three configurations on the left, produce an output equivalent to the same, common reparametrization of the loop phase $\phi \rightarrow \gamma(\phi)$, with $\gamma$ given by Eq. (\ref{eq:gamma1}). With lossless components, all energy entering the device must exit from the same port, so the net scattering action is the acquisition of a phase factor $e^{i\gamma(\phi)}$. These elementary ``phase amplifiers'' can thus be viewed as a phase-controlled source of optical phase. The configuration on the right is a traditional Sagnac interferometer, and only produces the phase factor of $e^{i\phi}$, unless a non-reciprocal phase is inserted in the loop. In that case, energy is split between the two output ports depending on the strength of the non-reciprocal phase.}
    \label{fig:configs}
\end{figure}

Perhaps the simplest example of a linear optical phase amplifier can be formed from an elementary optical ring cavity. As shown in Fig. \ref{fig:configs} (left), this could be generated by looping together two ports of a beam-splitter, if the two ports lie on opposite facets or share the same surface of reflection. The other configuration produces a traditional Sagnac interferometer and will not be used here. An analogous implementation of the looped-splitter ring cavity is also formed by coupling a microring resonator to a single bus waveguide. In either case, we assume the loop of length $\ell$ carries a physical phase $\phi = 2\pi n(\lambda) \ell/\lambda$, where $\lambda$ is the wavelength of a perfectly coherent monochromatic source and $n$ is the refractive index of the material forming the loop. Then the output amplitude $b$ is found by summing all possible paths through the system, which are indexed by the number of round-trips taken through the ring cavity. For the case where the looped facets are adjacent to one another, one finds \cite{PhysRevA.109.053508}
\begin{equation}\label{eq:b}
    b = \bigg ( r + t^2 e^{i\phi} \sum_{N = 0}^{\infty} (re^{i\phi})^N \bigg ) = r + \frac{t^2 e^{i\phi}}{1 - r e^{i\phi}} = \frac{r - (r^2 - t^2) e^{i\phi}}{1 - re^{i\phi}},
\end{equation}
where $r$ and $t$ are the reflection and transmission coefficients of the beam splitter, respectively. If the beam-splitter is lossless, then $|b| = 1$ regardless of the particular values of $r$ and $t$. This allows $b$ to be expressed as a phase factor, i.e. $b = e^{i\gamma(\phi, r, t)}$ for some function $\gamma$. This fact underlies the rest of this article, so we will now establish it in greater detail, after which we conclude the section by obtaining the particular functional form of $\gamma$ for this looped beam-splitter configuration. 

We first note that, being lossless, the beam-splitter must have a unitary scattering matrix. This implies both $|r|^2 + |t|^2 = 1$ and $r^*t + t^*r = 0$. Then, defining $r = |r|e^{i\phi_r}$ and $t = |t|e^{i\phi_t}$, the second constraint implies $\cos(\phi_r - \phi_t) = 0$, or that $\phi_r - \phi_t = \pi/2 + \pi n$ for any integer $n$. 

Next we consider the quantity $r^2 - t^2$, which appears in the final form of Eq. (\ref{eq:b}). We have 
\begin{equation}
    r^2 - t^2 = |r|^2 e^{2i\phi_r} - |t|^2 e^{i2\phi_t} = |r|^2 e^{2i\phi_r} - (1 - |r|^2) e^{i2\phi_r - i\pi - 2i\pi n} = e^{i2\phi_r}.
\end{equation}
Now, after placing this and the definition of $r = |r|e^{i\phi_r}$ into Eq. (\ref{eq:b}), $b$ becomes 
\begin{equation}\label{eq:b2}
b = \frac{|r|e^{i\phi_r} - e^{2i\phi_r + i\phi}}{1 - |r|e^{i\phi_r + i\phi}} = e^{i\phi_r} \bigg ( \frac{|r| - e^{i\phi_r + i\phi}}{1 - |r|e^{i\phi_r + i\phi}} \bigg )
\end{equation}
from which we can explicitly compute
\begin{equation}
    |b|^2 = b^* b = \bigg |e^{i\phi_r} \bigg ( \frac{|r| - e^{i\phi_r + i\phi}}{1 - |r|e^{i\phi_r + i\phi}} \bigg )\bigg |^2 = \frac{|r|^2 + 1 - |r| (e^{-i\phi_r - i\phi} + e^{i\phi_r + i\phi}) }{|r|^2 + 1 - |r| (e^{-i\phi_r - i\phi} + e^{i\phi_r + i\phi})} = 1.
\end{equation}
The value of $\phi_r$ bears no physical meaning, since it can be expressed as a global phase shift. This can be seen by writing out the full scattering matrix for this unitary beam-splitter, given by
\begin{equation}\label{eq:bsmatrix}
\begin{pmatrix}
  |r|e^{i\phi_r} & |t|e^{i\phi_t}\\
  |t|e^{i\phi_t} & |r|e^{i\phi_r}  
\end{pmatrix}
= 
\begin{pmatrix}
  |r|e^{i\phi_r} & -i e^{i\phi_r - i\pi n} \sqrt{1 - |r|^2}\\
  -i e^{i\phi_r - i\pi n} \sqrt{1 - |r|^2} & |r|e^{i\phi_r}  
\end{pmatrix}
 = e^{i\phi_r}
 \begin{pmatrix}
  |r|& \pm i \sqrt{1 - |r|^2}\\
  \pm i \sqrt{1 - |r|^2} & |r|
\end{pmatrix}.
\end{equation}
In tandem, we see the role of $\phi_r$ in Eq. (\ref{eq:b2}) is merely a fixed translation in the total phase of $b$ as well as a constant shift in the definition of $\phi$, both of which are globally indeterminate. 

So, again without any loss of generality, $\phi_r$ can be freely fixed at any value. We will take $\phi_r = 0$, so that $r$ is real ($r = |r|$) and $r^2 - t^2 = 1$. \textit{Within this choice of gauge, then, we may simplify Eq. (\ref{eq:b2}) to} 
\begin{equation}\label{eq:b3}
   b = \frac{r - e^{i\phi}}{1 - re^{i\phi}}.
\end{equation}
Now defining $\gamma = \arg b$, so that $b = e^{i\gamma}$, analysis of the preceding equation shows
\begin{equation}\label{eq:gamma1}
    \gamma(\phi, r) = \arctan \bigg (\frac{\sin\phi (1-r^2)}{\cos\phi(1+r^2) - 2r}\bigg)
\end{equation}
where $r \in [0, 1)$. In passing, we remark that in the third case of Fig. \ref{fig:configs}, where the looped facets are on opposite sides of the beam-splitter, the only change in the analysis of $b$ is that $r$ and $t$ are interchanged in Eq. (\ref{eq:b}). To obtain the analogous form of $\gamma$ in Eq. (\ref{eq:gamma1}), but parametrized in terms of a real-valued $t$ instead of a real-valued $r$, the gauge $\phi_r = \pi/2$ must be employed.

In general, the beam-splitter coefficients $r$ and $t$ are readily made phase tunable. This could be done for example by replacing the static beam-splitter with a constant-phase Mach-Zehnder interferometer \cite{Yang2024} or a standard Mach-Zehnder operated in the dual-drive configuration, where the two arms are locked to oppositely directed phases about a common bias point. Denoting $\delta$ as the phase which tunes the splitting ratio of the beam-splitter, we may write $r = \cos\delta$ (and $t = i\sin\delta$) for a phase $\delta$. Comparing this parametrization of $r$ and $t$ in terms of $\delta$ to Eq. (\ref{eq:bsmatrix}), we see it is expressed in the gauge employed in Eqs. (\ref{eq:b3})-(\ref{eq:gamma1}). In this gauge, then $\gamma$ becomes 
\begin{equation}\label{eq:gamma2}
    \gamma(\phi, \delta) = \arctan \bigg (\frac{\sin(\phi)\sin^2(\delta)}{\cos(\phi)(1 + \cos^2(\delta)) - 2\cos(\delta) } \bigg ).
\end{equation}
This is a common parametrization present in many physically distinct devices. The slight modification of adding $\phi$ to the above $\gamma$ would produce the parametrization realized by a Grover-Sagnac interferometer \cite{Schwarze:25}, in which $\delta$ is a non-reciprocal phase. This added $\phi$ may be lumped into the definition of the reference phase used to read out $\gamma$, so it can be discarded without loss of generality. The same phase response $\gamma$ may also be produced in a fully-reflective system by forming a Gires-Tournois interferometer with tunable reflectance. This is readily accomplished by adding a retro-reflecting mirror to one of the two ports of a Grover-Sagnac interferometer.

Although an infinite number of different phase parametrizations exist, \textit{the form of Eq. (\ref{eq:gamma2}) will be the working example for the remainder of this work}. Similarly we will only consider using a traditional Mach-Zehnder interferometer as a readout for the phase-amplified $\gamma$, although other phase readout configurations might be considered. Assume $\gamma$ is applied in one arm and in the other arm a reference phase $\theta$ is placed. The output reflection or transmission probability $P$ for single-photon input, which is proportional to the output power for coherent input, will still vary continuously between 0 and 1. However, the behavior of $P$ in a neighborhood around some point $\phi_0$ will depend strongly on $\gamma$. In particular, 
\begin{equation}\label{eq:P}
    P(\gamma(\phi, \delta), \theta) = \sin^2((\gamma(\phi,\delta) - \theta)/2).
\end{equation}
If $\gamma$ could be controlled directly, sweeping $\gamma$ would produce a standard sinusoidal interferogram. But since only the physical phase $\phi$ can be directly controlled, the interferogram $P$ vs. $\phi$ will become altered according to how $\gamma$ reparametrizes $\phi$. This can locally create regions with relatively high or low slope, curvature, etc. compared to the original parametrization which is implemented with a traditional phase shift $\phi$ in place of $\gamma(\phi)$. These different regions then form the bases of different computational operations.

\section{Perturbative phase optical computing}

In the following we consider various series expansions which allow $P(\gamma(\phi, \delta), \theta)$ to act as a surrogate for a desired function $L(\phi, \delta, \theta)$. $L$ is the quantity we would like to compute, and $P$ is what the optical system produces in response to the varying phase inputs. 

\textit{P} is the probability that an incident photon emerges from a given port of an interferometer. The collection of probabilities corresponding to all of the output ports forms a discrete probability distribution. The values of the underlying probabilities and therefore the aggregate distribution vary as the values of the phase parameters change. In other words, the internal phases parametrize a family of discrete probability distributions, and it is the nature of this parametrization which will be used to approximate computational operations. In this work, the focus is confined to a single output port's probability.

Although the computational signal is measured with a scattering probability measurement, the underlying mathematical operations are encoded in only the phase portion of the complex-valued scattering amplitude. The reference phase at the readout is always biased to ensure the transition from this phase-encoded amplitude to probability is as direct as possible at the bias point. In the present context, where all realized operations are scalar-valued, this encoding simplifies the process of tailoring computations out of a phase reparametrization $\gamma$ and enacting them directly with phase shifters.

As with other perturbative expansions, the error associated with using $P$ in place of $L$ can be made smaller by moving closer to the expansion point. In the first and simplest case, optical phase addition will be read out by a change in optical power proportional to the sum of the changes in power which would have resulted if they were applied individually. As a result, we have the computation $L(\phi, \delta, \theta) = \phi + \theta$ being well approximated by $P(\gamma(\phi, \delta), \theta)$ for a fixed $\delta$ and variable $\phi, \theta$ in some neighborhood of a bias point $(\phi_0, \theta_0)$.

This perturbative approach to addition in a linear optical system is analogous to employing a Hookean spring for the addition of numbers which are encoded by the masses of material bodies. Assuming the spring constant is known (meaning the scale is calibrated), the mass inputs are verified from the individual displacements they produce when affixed to the spring under the influence of gravity. Then the \textit{sum} of two or more masses is encoded in the total displacement of the spring when all are hung from the scale simultaneously. Hence the scheme requires linearity, assuming the sum of forces produces a displacement which is equal to the sum of the displacements which would have resulted from individual application of each force.

After addition, which naturally includes the inverse operation subtraction, we employ the coupling between $\phi$ and $\delta$ to design a proxy for multiplication as well as the nonlinear function $L(\phi, \delta) = \delta^2/\phi$. This coupling may be represented by a non-vanishing term $\partial^2 P/\partial \phi \partial \delta$ which is absent from traditional interferometers. Then we end with a discussion of cascaded scalar operations.

\subsection{Addition and subtraction}
First consider the expansion of $P$ (Eq. (\ref{eq:P})) about a fixed bias point $(\phi_0, \theta_0)$. This expansion and thus all of its coefficients are functions of the variable $\delta$. To leading order in perturbations $\Delta \phi$ and $\Delta \theta$, we have 
\begin{subequations}
\begin{align}\label{eq:expansion}
    &P(\gamma(\phi_0 + \Delta \phi, \delta), \theta_0 + \Delta \theta) \approx \\\notag &P(\gamma(\phi_0, \delta), \theta_0) + \frac{\partial P}{\partial \gamma}\frac{\partial \gamma}{\partial \phi}\bigg |_{(\phi_0, \delta, \theta_0)} \Delta \phi + \frac{\partial P}{\partial \theta}\bigg |_{(\phi_0, \delta, \theta_0)} \Delta \theta = \\ 
    &\sin^2((\gamma(\phi_0, \delta) - \theta_0)/2) + \frac12 \sin(\gamma(\phi_0, \delta) - \theta_0) (\gamma'(\phi_0, \delta) \Delta \phi - \Delta\theta), \label{eq:expansionfull}
\end{align}
\end{subequations}
where $\gamma'(\phi_0, \delta)$ is being used as shorthand for the function $\partial \gamma/\partial \phi$, evaluated at the bias point. Now set the bias phases to satisfy $(\gamma(\phi_0, \delta) - \theta_0)/2 =\pi/4 + 2\pi n$ for any integer $n$. For the $\gamma$ in Eq. (\ref{eq:gamma2}), $\gamma = 0$ for any value of $\delta$ when $\phi = 0$. So without loss of generality we select $\phi_0 = 0$ and $\theta_0 = 3\pi/2$. Then $\sin^2((\gamma(\phi_0, \delta) - \theta_0)/2) = 1/2$ and $\sin(\gamma(\phi_0, \delta) - \theta_0) = 1$, so the expansion reduces to 
\begin{equation}
    P \approx \frac12 + \frac12 (\gamma'(\phi_0, \delta) \Delta \phi - \Delta \theta).
\end{equation}
Relative to this bias point, the change in probability $\Delta P$ is given by
\begin{equation}\label{eq:deltaP}
    \Delta P = \frac12 (\gamma'(\phi_0, \delta) \Delta \phi - \Delta \theta).
\end{equation}
Thus, by measuring this change in probability we subtract the input phase perturbations. The perturbation $\phi$ is also scaled by the phase amplification factor from the parametrization $\gamma$, allowing generic linear transformations of $\Delta \phi$ to be obtained, up to an overall scale factor. The reference arm could readily be given a parametrization of its own as well, but for now we set $\delta = \pi/2$ so that $\gamma(\phi) = \phi$, $P$ is sinusoidal, $\gamma'(\phi_0, \delta) = 1$, and $\Delta P = (\Delta \phi - \Delta \theta)/2$.

$\Delta P$ is a signed quantity. The input perturbations can readily be directed with opposite signs, generating the addition of $\Delta \phi$ and $\Delta \theta$. Or, we might set $\Delta \theta = 0$ and apply two perturbations $\phi_1$ and $\phi_2$ in series in the same arm to naturally sum the phases, leading to $\Delta P = (\Delta \phi_1 + \Delta \phi_2)/2$, which again can be directed with opposite signs in order to subtract. This notion can be applied more than once: a series of separately controlled phase shifters in the same arm will all be summed together \cite{10.1063/1.5126517, d140ec05d7604fad9e94354f7758a18e}. 

When $\delta$ is shifted away from $\pi/2$, the strength of the summed phase perturbation in that arm is modified according to the phase amplification factor. For $\gamma(\phi_0, \delta) > 1$, the output signal is amplified, making detection easier. However, the dynamic range is reduced due to an increase in magnitude of the higher-order expansion coefficients at bias. Conversely, for $\gamma(\phi_0, \delta) < 1$, the size of the interval $(-\Delta\phi, \Delta \phi)$ over which addition can be conducted below a given error threshold can be substantially increased. This is similarly due to a decrease in the magnitude of higher-order coefficients which accompanies the decrease in slope at bias. 

\begin{figure}
    \centering
    \includegraphics[width=0.75\linewidth]{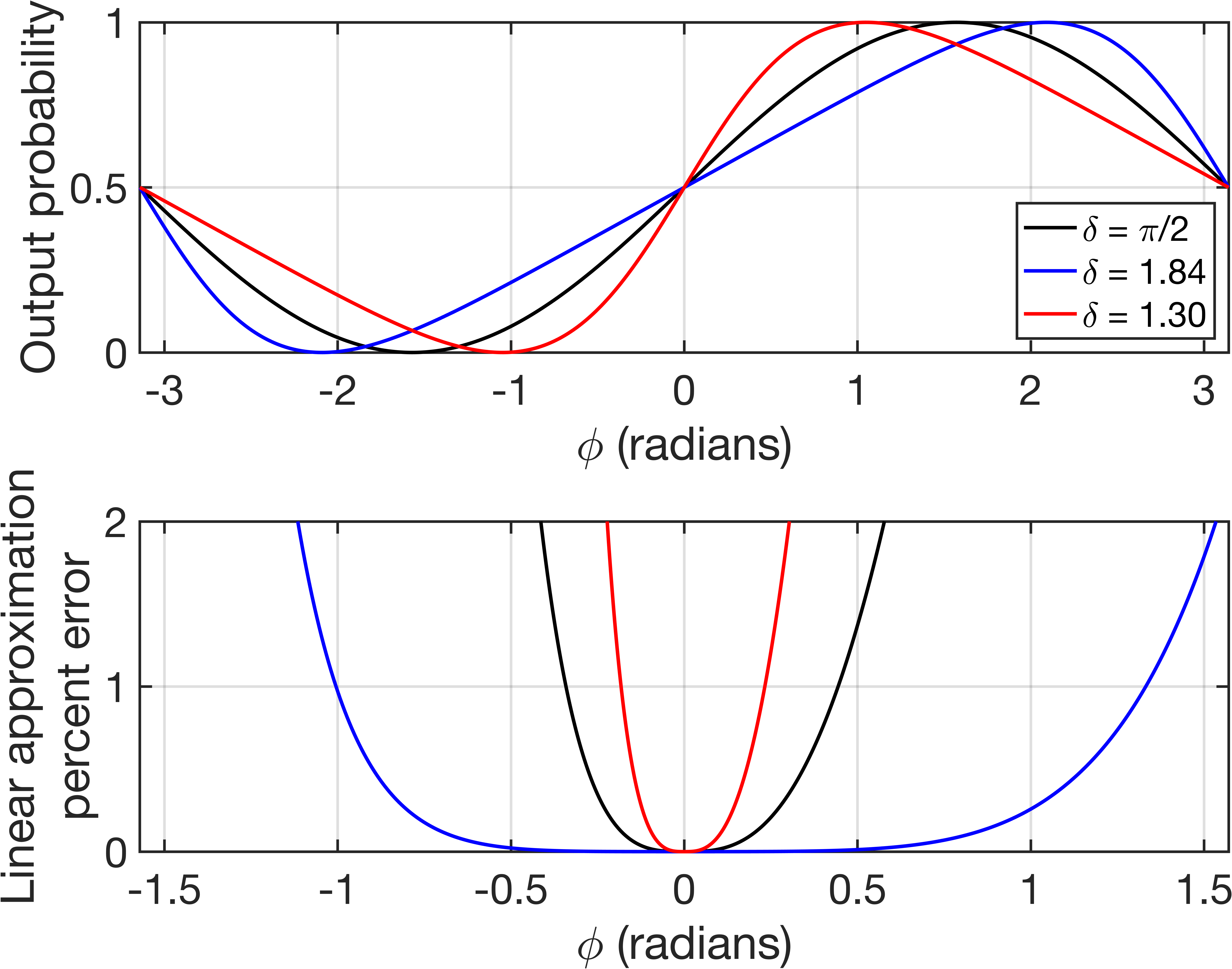}
    \caption{(top) Output probability $P(\gamma(\phi, \delta))$ vs. phase $\phi$ of a lossless, phase-amplified Mach-Zehnder interferometer for various phases $\delta$. $P = \sin^2((\gamma(\phi, \delta) - 3\pi/2)$ is the probability of a single photon entering a given input port emerging at one of the two output ports, with that of the other output port being $1 - P$. The particular amplification function $\gamma$ used here is defined in Eq. (\ref{eq:gamma2}). (bottom) Percent error in treating $P$ as a linear function about the origin for the same $\delta$ in the above case. If shifted by the value $-1/2$, the target function and approximant would be odd functions about $\phi = 0$. This implies that the absolute error is an even function about $\phi = 0$, since the mutual offset cancels there. However, the offset in the target function remains in the calculation of relative (percent) error, breaking that symmetry, as evident in the plot.}
    \label{fig:addition}
\end{figure}

Examples of these different operational regimes are shown in the top portion of Fig. (\ref{fig:addition}). The percent error associated with using these functions as an approximation to a line, upon which phases may be summed exactly, is shown in the lower portion. This error is only generated in the phase interval around the origin where the true, linear function is greater than zero to avoid singularities present in the definition of percent error.

The linearity assumption will begin to break down as the displacements from the bias point become large. The addition is generally valid to leading order (in $\Delta \phi_j$), as we are using a nonlinear sinusoid as an approximate surrogate for the ideal linear addition function. $\theta_0 = 3\pi/2$ was selected deliberately for the sake of biasing at an inflection point, where the second derivative of $P$ with respect to $\phi$ vanishes at bias. This comparatively improves the accuracy at larger displacements.

Outside the present context, this regime is generally able to improve the overall linearity of a Mach-Zehnder interferometer's response. Schemes where this may beneficial include optical phase feedback systems, phase imaging systems, and other computational schemes \cite{Ahmed2025}. Or, with a single phase perturbation $\Delta \phi$, varying $\delta$ allows $\Delta \phi$ to be rescaled by a constant. 

Effective addition and subtraction can be obtained by selecting some $\Delta \phi_{\text{max}}$ which bounds the error to a desired maximum. The perturbations are made with the quantity $\epsilon_j\Delta \phi_{\text{max}}$ for continuously varying $\epsilon \in [-1, 1]$. This is equivalent to tuning the continuously varying parameter $\phi$ in the set $(\phi_0 - \Delta\phi_{\text{max}},\phi_0 + \Delta\phi_{\text{max}})$. Selecting a smaller $\Delta \phi_{\text{max}}$ makes the procedure more accurate, but makes the changes in power smaller and therefore intrinsically more difficult to resolve. It also imposes tighter restrictions on the granularity or resolution of $\epsilon_j$.

The variables $\epsilon_j$ are of arbitrary units. Units are in effect chosen by deciding, during the act of decoding, how a \textit{single} input phase affects output probability change. One can envision many schemes for this, which may depend on how one desires to integrate this computing unit in a larger system. If quantified by percent error, the computation accuracy is independent of the choice of units. 

\subsection{Multiplication}
As a function of both $\phi$ and $\delta$, $\gamma$ may be viewed as a $\delta$-indexed family of curves $\gamma_\delta(\phi)$ where the slope with respect to $\phi$ about some point is controlled by $\delta$. Locally the influence of $\delta$ on the slope of $\gamma$ with respect to $\phi$ is described by a nonvanishing mixture term 
\begin{equation}\label{eq:crossp}
\frac{\partial}{\partial\delta}\bigg (\frac{\partial\gamma}{\partial\phi}\bigg) =\frac{\partial^2\gamma}{\partial\delta \partial\phi}
\end{equation}
in the power series for $\gamma$ expanded about $(\phi_0, \delta_0)$. We may view the bias point $\delta_0$ as selecting a curve from the family. Perturbations $\Delta \delta$ away from $\delta_0$ perturb the curve itself, which causes a change the slope at the point $\phi_0$. If $\gamma$ were linear in its arguments, this term would vanish, as is the case for a traditional Michelson or Mach-Zehnder interferometer.

The cross-partial term (\ref{eq:crossp}) describes first-order coupling between $\phi$ and $\delta$ and becomes relevant once \textit{both} $\phi$ and $\delta$ are perturbed. Carrying out this latter expansion, we obtain,
\begin{subequations}
\begin{align}\label{eq:expansion2}
    P(\gamma(\phi_0 + \Delta \phi, \delta_0 + \Delta \delta), \theta ) &\approx P(\gamma(\phi_0, \delta_0), \theta) + \frac{\partial P}{\partial \phi}\bigg |_{(\phi_0, \delta_0, \theta)} \Delta \phi + \frac{\partial P}{\partial \delta}\bigg |_{(\phi_0, \delta_0, \theta)} \Delta \delta + \frac{\partial^2 P}{\partial \phi \partial \delta}\bigg|_{(\phi_0, \delta_0, \theta)} \Delta\phi\Delta\delta \\\label{eq:expansion3}
    &= \sin^2((\gamma(\phi_0, \delta_0) - \theta)/2) + \frac12 \sin(\gamma(\phi_0, \delta_0) - \theta) \bigg ( \frac{\partial\gamma}{\partial \phi}\bigg|_{(\phi_0, \delta_0)} \Delta \phi + \frac{\partial\gamma}{\partial \delta}\bigg|_{(\phi_0, \delta_0)}\Delta\delta + \\\notag &\frac{\partial^2\gamma}{\partial\phi\partial\delta}\bigg|_{(\phi_0, \delta_0)} \Delta \phi \Delta \delta \bigg ) + \frac12 \cos(\gamma(\phi_0, \delta_0) - \theta)\frac{\partial\gamma}{\partial \phi}\bigg|_{(\phi_0, \delta_0)}\frac{\partial\gamma}{\partial \delta}\bigg|_{(\phi_0, \delta_0)}\Delta\phi\Delta\delta.
\end{align}
\end{subequations}
Generally, alteration of the entire curve $\gamma_\delta(\phi)$ which results from $\Delta\delta$ will affect not just the slope, but also the offset and potentially higher-order terms at $\phi_0$. To attain leading order accuracy, the higher-order terms can be ignored, but a change in offset or so-called bias shift resulting from the $\partial \gamma/\partial \delta$ term will mask the change due to the product term proportional to $\Delta\phi\Delta\delta$. To observe the product of phase perturbations in isolation, this term must be removed, but this is readily accomplished. In fact, for this particular $\gamma$, $\partial^n \gamma/\partial \delta^n = 0$ for all $n$ when evaluated at $\phi = 0$, no matter the value of $\delta$. Similarly, setting $\theta = 3\pi/2$ will ensure that $\sin(\gamma(\phi = 0, \delta) - \theta) = 1$ and $\cos(\gamma(\phi = 0, \delta) - \theta) = 0$. This simplifies the expansion (\ref{eq:expansion3}) to
\begin{align}\label{eq:expansion4}
    P(\gamma(\phi_0 + \Delta \phi, \delta_0 + \Delta \delta), \theta ) &\approx \frac12 + \frac12 \bigg ( \frac{\partial\gamma}{\partial \phi}\bigg|_{(\phi_0, \delta_0)} \Delta \phi + \frac{\partial^2\gamma}{\partial\phi\partial\delta}\bigg|_{(\phi_0, \delta_0)} \Delta \phi \Delta \delta \bigg ), 
\end{align}
from which we see 
\begin{align}\label{eq:deltaP2}
\Delta P &\coloneqq P(\gamma(\phi_0 + \Delta \phi, \delta_0 + \Delta \delta), \theta = 3\pi/2) - P(\gamma(\phi_0 + \Delta \phi, \delta_0), \theta = 3\pi/2) \approx \frac12 \frac{\partial^2\gamma}{\partial\phi\partial\delta}\bigg|_{(\phi_0, \delta_0)} \Delta \phi \Delta \delta.
\end{align}
In this configuration, the reference signal is captured after $\phi$ is perturbed away from $\phi_0 = 0$. From this reference point, the change in $P$ which results from perturbing $\delta$ is proportional to the product $\Delta\phi\Delta\delta$.

\subsubsection{Finite difference approximations}

Additional analysis of the previous equation reveals its limitations and points toward a simple change which will substantially improve its accuracy. For this sake, we may conceptually treat $\Delta P$ as a homotopy in $\Delta \phi$ which is indexed by $\Delta \delta$. For each value of $\Delta \delta$, the homotopy function is approximately a line with slope which is approximately given by $(1/2) \partial^2 \gamma/\partial\phi\partial\delta|_{(\phi_0, \delta_0)} \Delta \delta$. The approximations are accurate for both small $\Delta \phi$ and $\Delta \delta$. 

From the previous subsection and exemplified by Fig. \ref{fig:addition}, it is known that the value of $\delta_0$ can be assigned to make the homotopy functions act linear in $\Delta \phi$ about $\phi_0 = 0$ over a broad region of $\Delta \phi$ and $\Delta \delta$. In the neighborhood of such a $\delta_0$, different values of $\Delta \delta$ produce a different slope $\partial(\Delta P)/\partial(\Delta\phi)$ at the origin $\Delta \phi = 0$. Therefore, the final aspect of producing an approximation proportional to the function $\Delta\phi\Delta\delta$ \textit{is maintaining the correct, constant variation of the slope} $\partial (\Delta P) /\partial (\Delta \phi)$ with respect to $\Delta \delta$. 

To assess this, we directly compute $\partial (\Delta P) /\partial (\Delta \phi)$ at the bias point $(\phi_0 = 0, \delta_0)$ as a function of $\Delta \delta$. This analysis is fundamentally tied to the underlying phase $\gamma$ and less so the probability, since the choice of $\theta$ and $\phi_0$ has established that $\partial P/\partial\phi|_{(\phi_0, \delta)} = (1/2) \partial \gamma/\partial\phi|_{(\phi_0, \delta)} = \gamma'(\phi_0, \delta)$ for any $\delta$. Differentiating Eq. (\ref{eq:deltaP2}) and substituting in this expression, we obtain
\begin{align}
    \frac{\partial (\Delta P)}{\partial (\Delta\phi)}\bigg|_{(\phi_0 = 0, \delta_0)} = \frac{\partial (\Delta P)}{\partial \phi}\bigg|_{(\phi_0 = 0, \delta_0)} = \frac12(\gamma'(\phi_0 = 0, \delta_0 + \Delta \delta) - \gamma'(\phi_0 = 0, \delta_0)). 
\end{align}
The right hand side can be identified as a \textit{forward} finite difference approximation for the derivative of $\gamma'$ with respect to $\delta$. According to such,
\begin{equation}
    \gamma'(\phi_0 = 0, \delta_0 + \Delta \delta) - \gamma'(\phi_0 = 0, \delta_0)\approx \frac{\partial \gamma'}{\partial \delta}\bigg|_{(\phi_0 = 0, \delta_0)} \Delta \delta = \frac{\partial^2\gamma}{\partial\phi\partial\delta}\bigg|_{(\phi_0 = 0, \delta_0)}\Delta \delta.
\end{equation}
The approximation in the preceding equation is valid to order $\Delta \delta$ but can be improved to order $\mathcal{O}(\Delta \delta)^2$ by employing to a central difference (CD) approximation. The corresponding probability difference is 
\begin{equation}\label{eq:deltaPCD}
\Delta P_{\text{CD}} \coloneqq P(\gamma(\phi_0 + \Delta \phi, \delta_0 + \Delta \delta/2), \theta = 3\pi/2) - P(\gamma(\phi_0 + \Delta \phi, \delta_0 - \Delta\delta/2), \theta = 3\pi/2).
\end{equation}
A backward difference could also be employed, but this provides no more accuracy than the forward difference. 

\begin{figure}[htb!]
    \centering
    \includegraphics[width=0.75\linewidth]{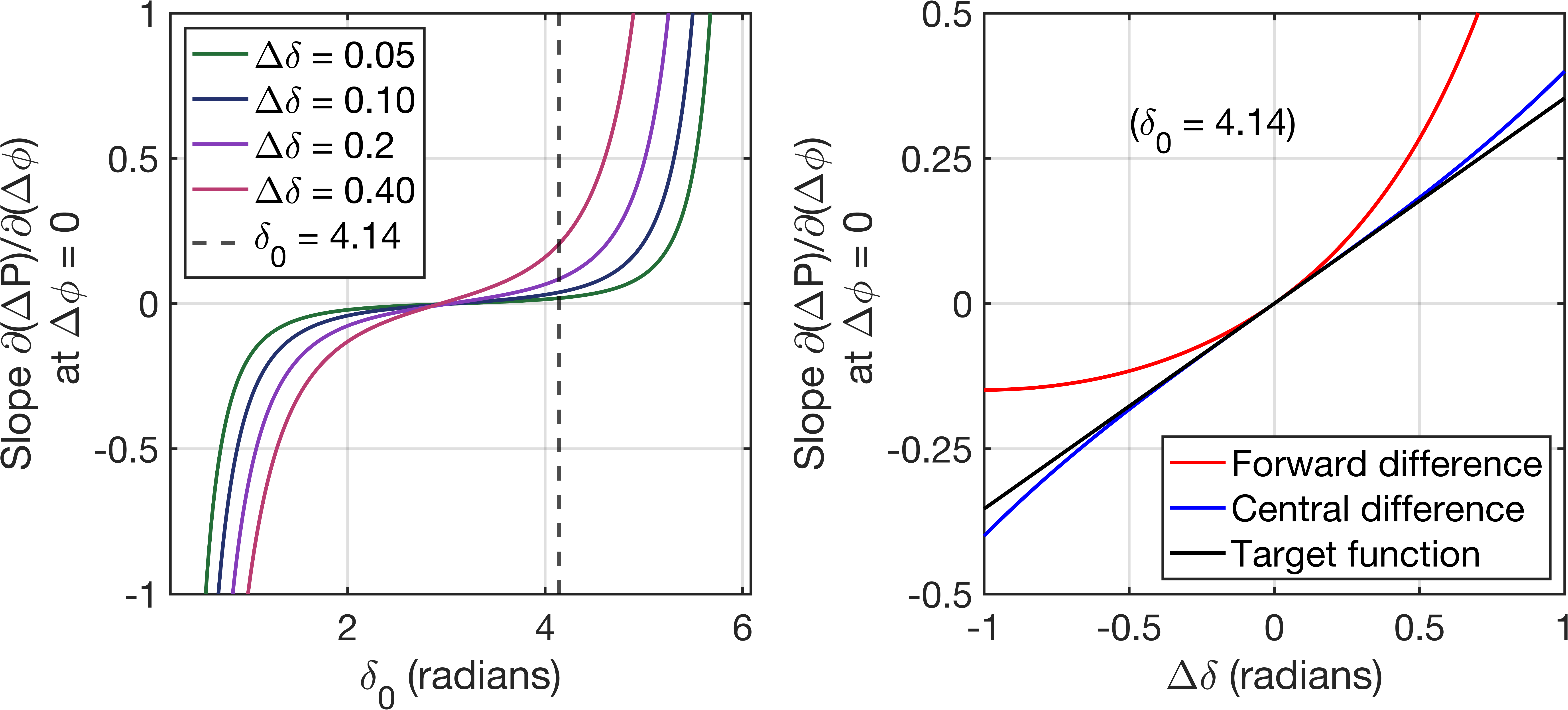}
    \caption{To produce a $\Delta P$ approximately proportional to the function $\Delta\phi\Delta\delta$, the slope of $\Delta P$ with respect to changes in $\Delta \phi$, evaluated at the bias point, should be proportional to $\Delta \delta$ and with nonzero proportionality constant. This slope is shown on the left for several values of $\delta_0$. Selecting $\delta_0$ to induce linear behavior in $\Delta \phi$, such as around $\delta_0 = 4.14$ (dashed line), leads to a comparatively nonlinear response in $\Delta \delta$ when it is ideally linear. The effect of this nonlinearity can be reduced with a central difference approximation for $\Delta P$ instead of the forward difference one. This is illustrated by the plot on the right, which compares the two approximations to the linear target function, $(1/2) \partial^2P/\partial\delta\partial\phi|_{(\phi_0, \delta_0)}\Delta \delta$ at the bias point $(\phi_0, \delta_0) = (0, 4.14)$.}
    \label{fig:centraldiff}
\end{figure}

Conceptually, the relative accuracy of the employed finite difference approximation reduces to the degree to which the function $\Delta P$ is linear in $\Delta\delta$ around $\delta_0$, after $\delta_0$ has been fixed in order to produce linear behavior in $\Delta\phi$ about $\phi_0 = 0$. For this particular $\gamma$, the change in $\Delta P$ with respect to $\Delta \phi$ is given by 
\begin{equation}
\frac{\partial (\Delta P)}{\partial (\Delta \phi)}\bigg|_{\Delta \phi = 0, \delta = \delta_0 + \Delta\delta} = \frac12 \bigg(\cot^2\bigg(\frac{\delta_0 + \Delta\delta}{2}\bigg) - \cot^2\bigg(\frac{\delta_0}{2}\bigg)\bigg)
\end{equation}
and is shown in Fig. \ref{fig:centraldiff} (left) for several values of $\Delta \delta$. Selecting a $\delta_0$ which induces linear behavior in $\Delta \phi$ (e.g. around the value $\delta_0 = 4.14$) will come at the cost of considerable curvature in the region around $\delta_0$ where $\Delta \delta$ is to be varied. As a result, the central difference approximation significantly outperforms the forward difference one here, as exemplified in Fig. \ref{fig:centraldiff} (right). The target function is obtained by evaluating the actual cross-partial derivative. For $\gamma$ in Eq. (\ref{eq:gamma2}), $\partial^2 \gamma/\partial\phi\partial\delta|_{(\phi_0=0, \delta_0)} = -\cot(\delta_0/2)\csc^2(\delta_0/2)$.

\begin{figure}[htb!]
    \centering
    \includegraphics[width=0.75\linewidth]{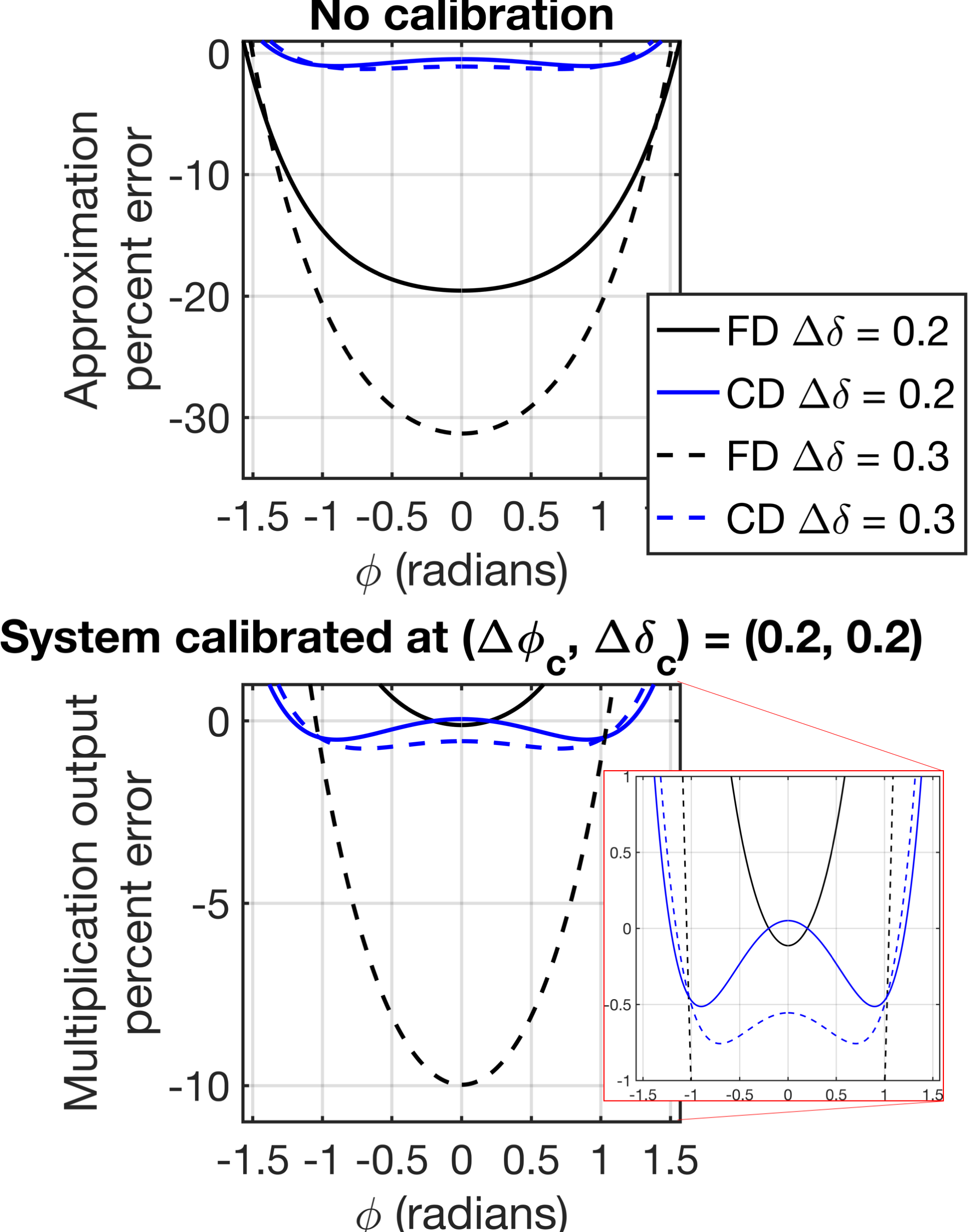}
    \caption{(Top) Percent error for multiplication approximation. The true function is $(1/2)\frac{\partial^2P}{\partial\delta\partial\phi}|_{(\phi_0, \delta_0)}\Delta\phi\Delta\delta$, the forward difference (FD) approximant is given by Eq. (\ref{eq:deltaP2}) and the central difference (CD) approximant is given by Eq. (\ref{eq:deltaPCD}). (Bottom) Percent error for calibrated multiplication. Calibration entails rescaling the approximant by the value which brings the error to zero at an arbitrarily chosen calibration point, here chosen to be $(\Delta\phi_c, \Delta\delta_c) = (0.2, 0.2)$. The central difference approximation is considerably more robust to changes in $\Delta\delta$ away from the calibration point. Because the target and approximant are both symmetric functions about the expansion point $\phi = 0$ (for all values of $\delta$), the computed error functions maintain this even symmetry.}
    \label{fig:multiplication}
    \end{figure}

A plot contrasting the approximation percent error for forward difference approximant of Eq. (\ref{eq:deltaP2}) and central difference approximant of Eq. (\ref{eq:deltaPCD}) is shown in Fig. \ref{fig:multiplication} (top). Again the system is biased at $(\phi_0, \delta_0, \theta_0) = (0, 4.14, 3\pi/2)$. Although the central difference approach is considerably more accurate, especially when $\Delta \delta$ is varied, neither are truly perfect when $\Delta\delta$ is greater than zero. Notwithstanding, a rudimentary calibration of the output signal can be conducted wherein the signal at a user-chosen calibration point $(\Delta\phi_c, \Delta\delta_c)$ is rescaled by a constant, such that the error is zero at the calibration point. This simple rescaling can e.g. be accomplished by tweaking the gain of the transimpedance amplifier of the photocurrent, or perhaps optically with an additional phase amplification stage. In any case, applying that constant scale factor to all other measurements further reduces the error around the neighborhood of the calibration point. This is exemplified in Fig (\ref{fig:multiplication}) (bottom), where $(\Delta\phi_c, \Delta\delta_c) = (0.2, 0.2)$. There it can be observed the error is zero at the calibration point and near zero around it as $\Delta \phi$ is varied. Moreover, the central difference approximant is considerably more robust to changes in $\Delta\delta$ away from the calibration point. 

It is possible a different interferometric configuration, possibly in conjunction with a more sophisticated, dynamic recalibration procedure and/or the use of higher-order finite difference approximations naturally admits a more accurate product. The current configuration is favorable due to its lack of bias shift, i.e. $\partial \gamma/\partial \delta$ is zero at bias as well as its relatively low resource count. Notwithstanding, an arbitrary $\gamma$ with a bias shift can still effectively be used for phase multiplication if the contributions for both individual perturbations are subtracted away. This method allows any bias point $(\phi_0, \delta_0)$ to be used, and defines $\Delta P \coloneqq P(\gamma(\phi_0 + \Delta \phi, \delta_0 + \Delta \delta), \theta) - P(\gamma(\phi_0 + \Delta \phi, \delta_0), \theta) -  P(\gamma(\phi_0, \delta_0 + \Delta \delta), \theta) + P(\gamma(\phi_0, \delta_0), \theta)$. The subtractions could either be done sequentially with multiple measurements or in parallel with multiple computation units. 

\subsection{Nonlinear operations}
In the prior subsections, linear arithmetic operations were obtained with approximately linear regions of the phase amplification map $\gamma$. Now we consider nonlinear regions of the $\gamma$ of Eq. (\ref{eq:gamma2}) to obtain a nonlinear approximant for scalar inversion which outperforms the associated linear approximation. Around the point $\phi = \delta$, the function $\gamma(\phi, \delta)$ approximates $f(\phi, \delta) = -\delta^2/\phi$. Or, by redefining $\phi \rightarrow -\phi$, an approximant for $\delta^2/\phi$ is obtained. This redefinition is equivalent to perturbing the phase about its bias point in the opposite direction. This could be accomplished by inverting the polarity in standard electro-optic platforms, using a material with negative electro-optic coefficient \cite{Yang2024-2} with regular polarity, or in thermo-optic phase shifters, selecting a material with negative thermo-optic coefficient \cite{WIECHMANN20096847}.

\begin{figure}[htb!]
    \centering
    \includegraphics[width=0.5\linewidth]{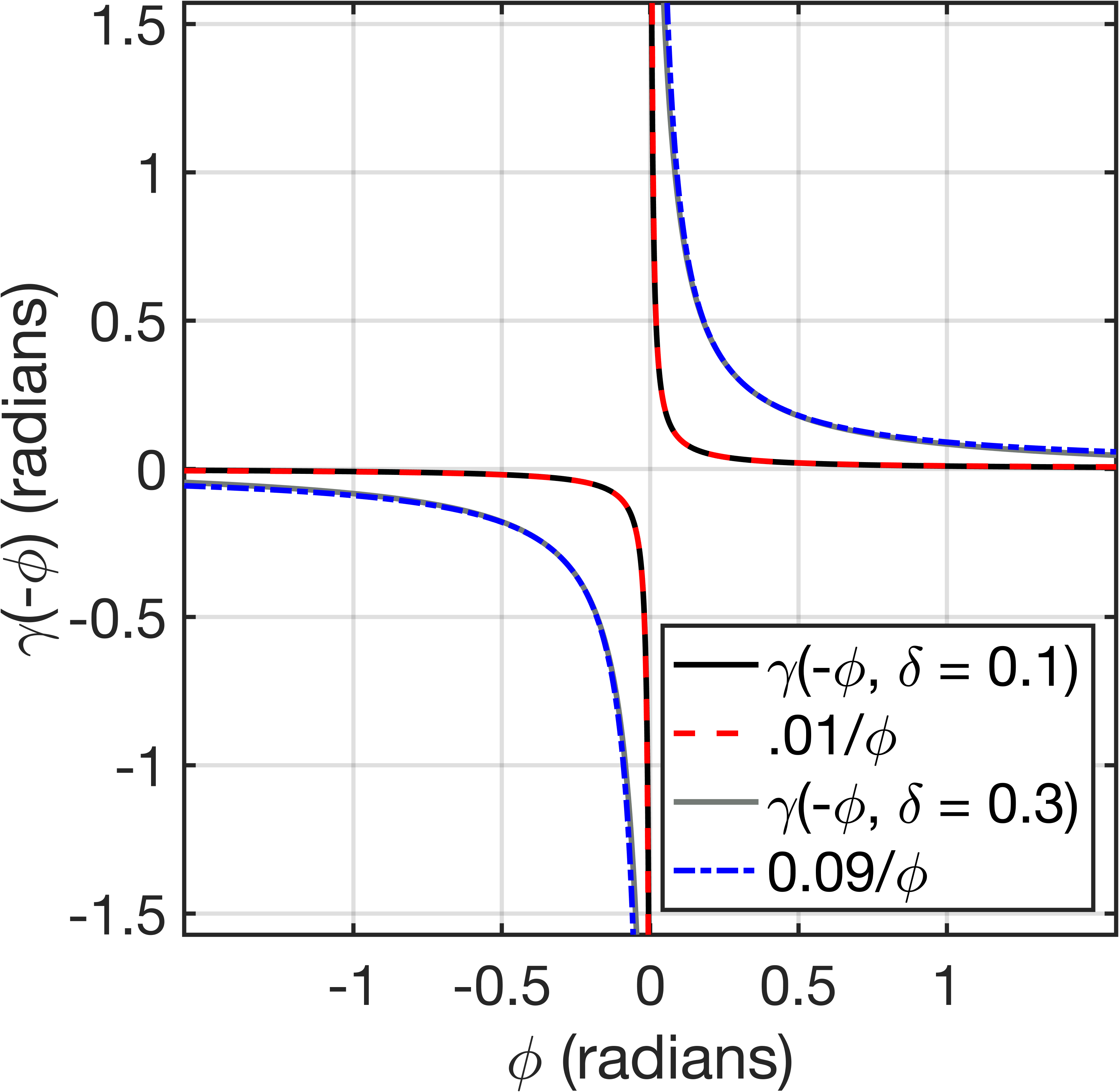}
    \caption{Two examples of $\gamma$ (Eq. \ref{eq:gamma2}, solid curves) operating around $\phi = \delta$ to approximate the function $\delta^2/\phi$ (dashed curves). The approximation breaks down near the origin, $\phi = 0$ and is $2\pi$ periodic. Zoomed-in details of these curves are presented in Fig. \ref{fig:division02}) and the associated probability is shown in Fig. \ref{fig:division03}.}
    \label{fig:division01}
\end{figure}

The validity of this approximation can be understood by comparing the leading order terms in the polynomial expansion of each function about the point $\phi = \delta$. We have 
\begin{subequations}\label{eq:inv}
\begin{align}
    \frac{\delta^2}{\phi} = \frac{\delta^2}{\delta + (\phi - \delta)} &\approx \delta - (\phi - \delta) +\frac{(\phi - \delta)^2}{\delta} - \frac{(\phi - \delta)^3}{\delta^2} + \mathcal{O}((\phi - \delta)^4), \\
    \gamma(-\phi, \delta) &\approx \delta - (\phi - \delta) +\frac{(\phi - \delta)^2}{\tan(\delta)}  -  \frac{(\phi - \delta)^3}{\tan^2 (\delta)} + \mathcal{O}((\phi - \delta)^4).
\end{align}
\end{subequations}
The expressions agree to first order, and remain similar through third order. The only difference is that $\delta$ is replaced with $\tan(\delta)$ in the denominator of the second and third order terms. These become more accurate as $\delta$ becomes smaller, since $\tan(\delta) \approx \delta + \mathcal{O}(\delta^3)$. At higher orders, however, the coefficients are less similar, and do not coalesce in the limit $\delta \rightarrow 0$. Examples of this $\gamma$ operating in this mode are shown in Fig. \ref{fig:division01}.

\begin{figure}[htb!]
    \centering
    \includegraphics[width=0.75\linewidth]{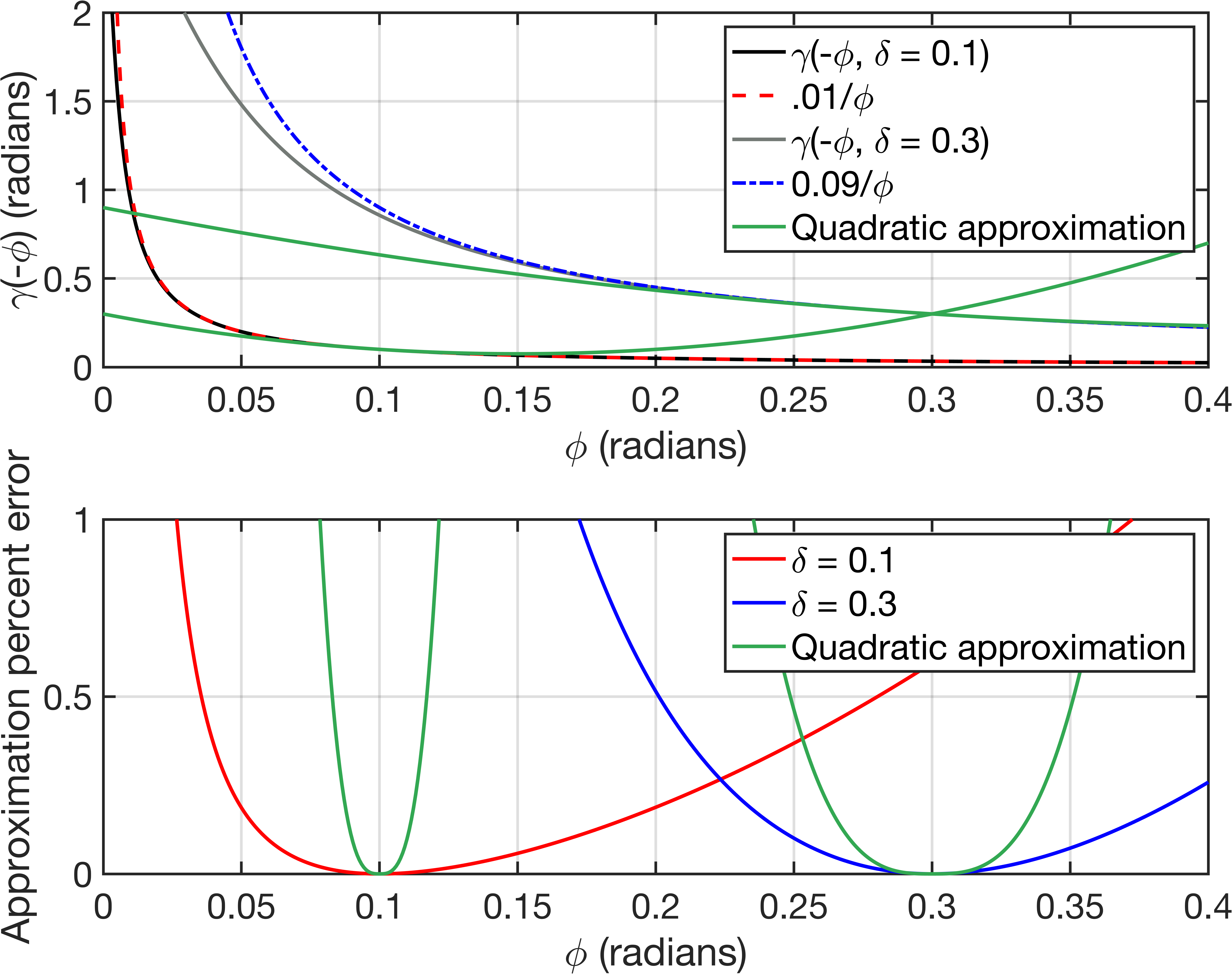}
    \caption{(top) Finer details of the curves shown in Fig. \ref{fig:division01}. Quadratic Taylor approximations are plotted for each case in green. (bottom) Associated percent error for each curve is shown. As $\delta$ decreases, the hyperbola $\delta^2/\phi$ sharpens and the range of quadratic approximation for a given accuracy margin shrinks. However, the associated dynamic range for $\gamma$ remains nearly constant with respect to $\delta$. The lack of symmetry in the error responses about each expansion point stems from the fact that the target function is not symmetric about these points.}
    \label{fig:division02}
\end{figure}

Although $\gamma$ is a nonlinear function, it is still, strictly speaking, only a first-order approximation to $\delta^2/\phi$ when the value of $\delta$ is unrestricted. For this more general case, as $\phi$ approaches $\delta$, a second-order Taylor expansion will eventually start to exhibit higher accuracy. However, $\gamma$ generally outperforms the regular linear first-order approximation, since its higher order coefficients better match those of the true function. Moreover, for larger values of $|\phi - \delta|$, $\gamma$ exhibits higher accuracy than even the second-order approximation, thereby exhibiting a higher-dynamic range for certain error thresholds. This is illustrated in the Fig. \ref{fig:division02}, where the curves of Fig. \ref{fig:division01} are shown at a finer scale in the top portion with the associated approximation errors shown below. 

As the error curves illustrate in Fig. \ref{fig:division02} (bottom), the dynamic range of this approximation $\gamma$ only weakly varies with $\delta$, whereas a traditional first-order approximation would exhibit a dynamic range strongly dependent on $\delta$. Intuitively, this is because $\gamma$ bends with the hyperbola $\delta^2/\phi$ as $\delta$ is varied. Beyond this, $\gamma$ induces lower error over a much larger range than a quadratic approximation.

The present form $\gamma \approx \pm \delta^2/\phi$ can be employed to obtain two elementary nonlinear functions, both of which are conic sections. For fixed $\phi$ it provides quadratic behavior in $\delta$, with the fixed $\phi$ value determining the parabola's scale or units. Likewise, fixing $\delta$ sets a scale for the scalar inversion of $\phi$, producing a hyperbolic response. In particular, after freely selecting $\delta_0$, $\phi$ is biased at $\phi_0 = \delta_0$. 

The nonlinear operating regime of $\gamma$ differs from the above cases in that $\phi$ is biased away from the origin at the value $\phi = \delta$. This shift in bias point generates a corresponding shift in the output probability or intensity when this $\gamma$ is read out. In general, when $\gamma$ is converted to a measurable quantity $P$ such as probability or optical intensity, it is composed with a function which may be viewed as a type of filter $f$. Ideally this filter is the identity map, since this would perfectly reproduce $\gamma$: $P = f(\gamma) = \gamma$ for all values of the underlying parameters. In reality there are theoretical limitations on the output of an interferometric readout, such as periodicity of the function, as well as experimental nuances which could alter the filter behavior, such as detector saturation. Therefore, in order to most faithfully reproduce a region of $\gamma$ at the final output, $\theta$ is selected so that an associated region of interest falls into a locally linear portion of the sinusoid.

In the two-beam overlap readout considered here, the filter is a sinusoidal function which can be translated by modifying the reference phase $\theta$. For $\theta = 3\pi/2$, we have 
\begin{equation}
    P = \sin^2((\gamma - 3\pi/2)/2) = \frac{1}{2} (1 + \sin(\gamma)) \approx \frac{1}{2} (1 + \gamma)
\end{equation}
where the approximation holds for small values of $\gamma(\phi, \delta)$, which is about the origin $\phi = 0$ for all $\delta$. However, in the present case, the region of interest is about $\phi = \delta$ for any given $\delta$. In order to make the approximation valid in an analogous way for the general case, $\theta$ must be shifted by the bias point to give
\begin{equation}\label{eq:nonzerobias}
    P = \sin^2((\gamma - \gamma_0 - 3\pi/2)/2) = \frac{1}{2} (1 + \sin(\gamma - \gamma_0)) \approx \frac{1}{2} (1 + \gamma - \gamma_0)
\end{equation}
where now the approximation holds for $\gamma$ near its bias point, i.e. small values of $|\gamma - \gamma_0|$. In the present case, we have
\begin{equation}
    P \approx \frac{1}{2} (1 + \gamma - \gamma_0) \approx \frac{1}{2} (1 + \frac{\delta^2}{\phi} - \delta),
\end{equation}
and so it can be seen that $P$ is now shifted by an additional term proportional to $\delta$ in addition to the standard offset of $1/2$. While decreasing $\delta$ can mitigate this shift, the shift could also be corrected in several ways, such as by the fan-in of an auxiliary beam calibrated to produce $\Delta P = \delta/2$. Incoherently summing the beams on the same detector will produce the desired result. Alternatively, two arbitrary computations can be added coherently by overlapping their beams at different ports of a beam-splitter. This is equivalent to nesting the individual computation circuits within different arms of a larger Mach-Zehnder interferometer. 

\begin{figure}[htb!]
    \centering
    \includegraphics[width=\linewidth]{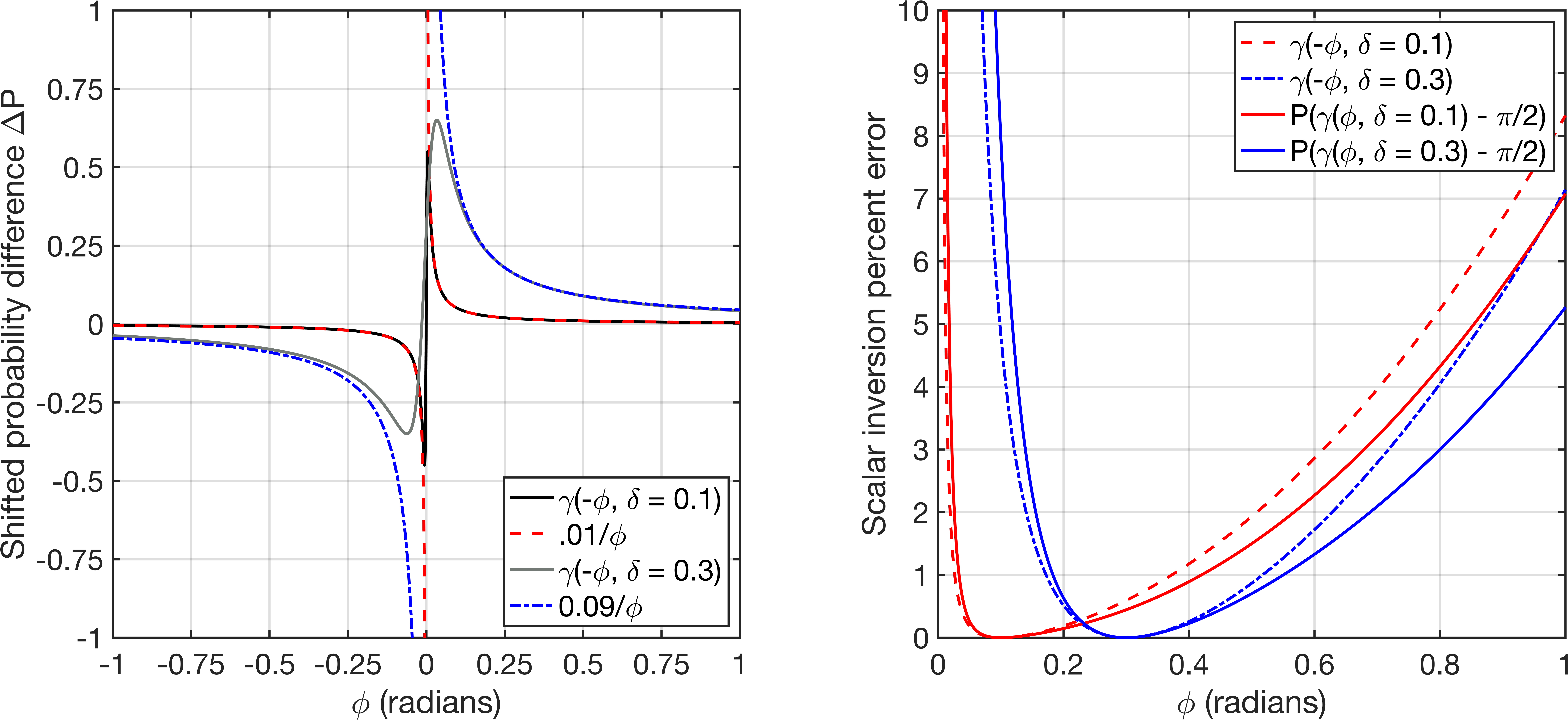}
    \caption{ (left) Outputted change in probability $\Delta P$ for $\gamma \approx \delta^2/\phi$. The probability has been shifted by an additional $\delta/2 - 1/2$ to produce $\Delta P$. The $\delta/2$ term is needed to correct a shift that results from being biased away from the origin $\phi = 0$, as discussed in the main text. Negative values of $\Delta P$ correspond to decreases in optical power while positive values correspond to increases in optical power. Although the target functions (blue and red curves) diverge as $\phi$ approaches the origin, $\Delta P$ (black and gray curves) must stay bounded. Therefore, once $\Delta P$ reaches its extreme values as $\phi$ is brought toward the origin, the approximation severely breaks down. (right) Corresponding percent error in approximation of scalar inversion using $\Delta P$ (solid curves) and $\gamma$ (dashed curves), illustrating that the phase approximation error is further modified by the readout. }
    \label{fig:division03}
\end{figure}

After using either of these procedures to shift the outputted probability difference from the scalar inversion circuit, a direct comparison to the desired function $\delta^2/\phi$ can be made. Examples of the output probability and its associated approximation error are shown in the Figure \ref{fig:division03}. The approximation breaks down near the origin $\phi = 0$ where the targeted hyperbola (blue and red curves in Fig. \ref{fig:division03}) approaches a vertical asymptote. The probability function (black and gray curves in Fig. \ref{fig:division03}) must be periodic and bounded, so it eventually must return through the origin $\Delta P = 0$ instead of continuing to diverge. At the point $(\phi, \delta) = (0, 0)$ there is a pole in $\gamma$. Moving toward this pole by further reducing $\delta$ improves the approximation in this region, causing the resonance about $\phi = 0$ to become increasingly narrow.

From Eq. (\ref{eq:nonzerobias}) we note adding an additional $\pi$ to $\theta$ would flip the sign of the $\sin$ function, allowing $-\gamma$ to be produced in the output probability instead. This eliminates the need to externally enact the map $\phi \rightarrow -\phi$ whenever $\gamma$ is being read out directly. However, this is not a general remedy when $\gamma$ is used within a larger phase-encoded procedure: there, the function $e^{-i\gamma}$ is needed, which is not the same as $e^{i\gamma + i\pi} = -e^{i\gamma}$.

\subsection{Cascaded scalar operations}
Cascading the phase parametrizations equates to a recursive nesting of their corresponding interferometers. This composes the phase functions they each produce, passing the output of one as input to the other. This is in contrast to methods which daisy-chain scattering devices in series, passing the output scattering state as an input state to the next device. 

\begin{figure}
    \centering
    \includegraphics[width=0.5\linewidth]{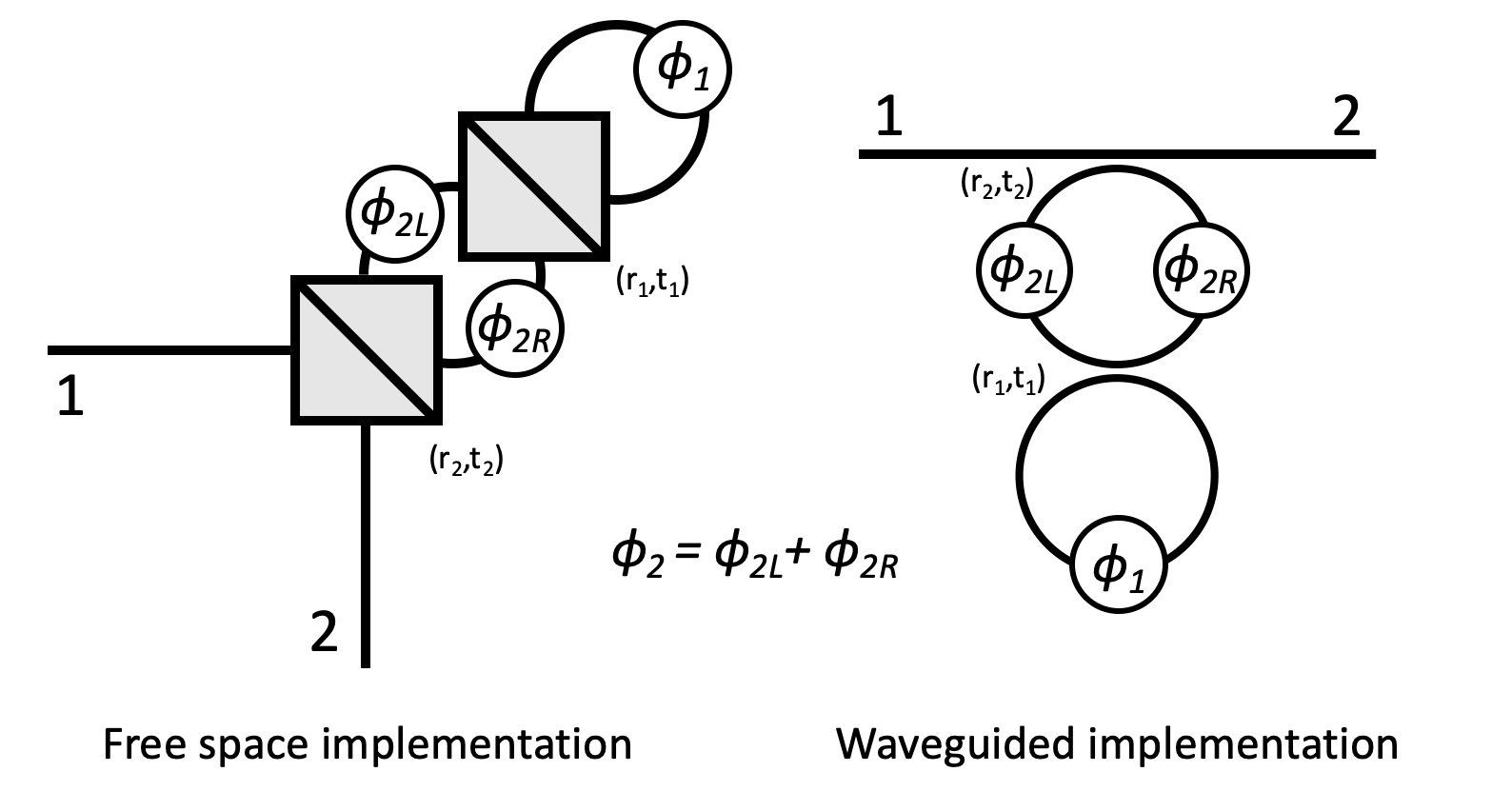}
    \caption{Nesting the interferometers will compose their underlying parametrizations, producing approximations for cascaded computations. Two equivalent cases are shown here, nesting the parametrization $\gamma$ of Eq. (\ref{eq:gamma2}) within itself, producing a final output phase factor of $\exp{i\gamma(\gamma(\phi_1, r_1) + \phi_{2L} + \phi_{2R}, r_2)}$. }
    \label{fig:rings}
\end{figure}

For example, it can be shown that placing the basic $\gamma$ unit of Fig. \ref{fig:configs} and Eq. (\ref{eq:gamma2}) inside the loop of \textit{another} similar unit would produce the aggregate phase transformation $\gamma(\gamma(\phi_1, r_1) + \phi_2, r_2)$, where $\phi_j$ is the total loop phase of ring $j$ and $r_j$ is the reflection amplitude for that ring's scattering interface. This configuration is shown schematically with tabletop beam-splitters and equivalently with guided ring resonators in Fig. \ref{fig:rings}. The portions of the outer ring's phase on either side of the inner unit need to be tracked separately, but the final output only depends on their sum $\phi_2$. 

There are multiple ways to establish that nesting these interferometers will compose their phase parametrizations. The following approach recursively uses of the single loop result of Eq. (\ref{eq:b}). Its validity is rooted in two facts. First, due to the feed-forward nature of this system, all of the light entering port 1 will exit at port 2. The final amplitude at port 2 is then the sum of all possible paths through the system, from port 1 to port 2, weighted by the scattering coefficients each path encounters as it traverses the two loops. The second fact is that this sum will be correct so long as all possible paths are accounted for, regardless of the manner of their enumeration. In particular, it makes no difference if combinatorially-derived subsets of paths are summed first, and then after this, these summed components are added together. This is because the overarching addition of all paths is commutative.

Exploiting this fact, one can immediately sum the round-trip paths that correspond to some amplitude portion entering the inner ring into a steady state, using the results for a single ring. The rest of the calculation is then reduced to counting the paths which enter the inner ring, scaling these by the corresponding inner-ring steady-state amplitude. The calculation proceeds as follows. When light enters the double loop system at port 1, a transient amplitude of $r_2$ goes to the output with an amplitude $t_2$ exciting ring 2. This acquires the $\phi_{2R}$ portion of the total phase around the ring $\phi_2$, and then couples to the other ring. A factor of $t_1$ is given to the portion coupling to ring 1 while $r_1$ goes to the portion staying in ring 2. If we identify photon creation operators $a_j^\dagger$ with the external port $j$ and likewise $b_j^\dagger$ with ring $j$, then the initial dynamics of the photon entering port 1 can be expressed like so:
\begin{equation}\label{eq:transient}
    a_1^\dagger \rightarrow r_2 a_2^\dagger + t_2 e^{i\phi_{2R}}(r_1 b_2^\dagger + t_1 b_1^\dagger).
\end{equation}
The recursion relation for $b_2^\dagger$ making one full round trip is 
\begin{equation}
    b_2^\dagger \rightarrow e^{i\phi_{2L}} t_2 a_2^\dagger + r_2 e^{i\phi_2}(r_1 b_2^\dagger + t_1 b_1^\dagger)
\end{equation}
From the analysis of the single loop system, it is known that in the steady state $b_1^\dagger \rightarrow \frac{t_1 e^{i\phi_1}}{1 - r_1 e^{i\phi_1}} b_2^\dagger$. This is precisely the intermediate summing of sub-paths described above. This result can be placed into the above formulae, which in turn reduces the $b_2^\dagger$ recursion formula to
\begin{equation}
    b_2^\dagger \rightarrow e^{i\phi_{2L}} t_2 a_2^\dagger + r_2 e^{i\phi_2}(r_1 + \frac{t_1^2 e^{i\phi_1}}{1 - r_1e^{i\phi_1}}) b_2^\dagger = e^{i\phi_{2L}} t_2 a_2^\dagger + r_2 e^{i\phi_2 + i\gamma_1(\phi_1)} b_2^\dagger
\end{equation}
so that the steady state mapping of $b_2^\dagger$ is
\begin{equation}
    b_2^\dagger \rightarrow \bigg ( \frac{t_2 e^{i\phi_{2L}}}{1 - r_2 e^{i\phi_2 + i\gamma_1(\phi_1)}} \bigg ) a_2^\dagger.
\end{equation}
By placing this result into the expression for $b_1^\dagger$'s steady state, we obtain $b_1^\dagger$'s steady state in terms of the output mode $a_2^\dagger$. Finally we place these two steady-state expressions (only in terms of $a_2^\dagger$) into the transient expression (\ref{eq:transient}), obtaining a final output amplitude $b$ of 
\begin{subequations}
\begin{align}
b &= r_2 + t_2 e^{i\phi_{2R}} \bigg ( \frac{t_2 e^{i\phi_{2L}}}{1 - r_2 e^{i\phi_2 + i\gamma_1(\phi_1)}} \bigg ) \bigg [ r_1  + t_1 \bigg (\frac{t_1 e^{i\phi_1}}{1 - r_1 e^{i\phi_1}} \bigg )  \bigg] \\ 
&= r_2 + \frac{t_2^2 e^{i\phi_2}}{1-r_2 e^{i\phi_2 + i\gamma_1(\phi_1)}} [e^{i\gamma_1(\phi_1)}] \\ &= 
e^{i\gamma_2(\phi_2 + \gamma_1(\phi_1))}.
\end{align}
\end{subequations}

A general source of error in cascading scalar functions arises if the output value of the inner function does not fall close to the center of the biased region of the outer function. In other words, the inner function needs to take on values near the outer function's bias point, or else the outer function's approximation is expected to break down. This issue is relevant with \textit{this specific $\gamma$} as it pertains to nesting inversion and multiplication to obtain direct divison. The multiplication regime is biased about $\phi = 0$ while inversion is biased about $\phi = \delta$. This means that for the outer multiplication to be valid, the output of the inversion must be near 0. The output of this inversion is approximately given by $\delta_1^2/\phi$ which itself is a good approximation when $\phi$ is near $\delta_1$. Increasing $\phi$ will improve the quality of the multiplication approximation at the expense of the inversion one.

\begin{figure}
    \centering
    \includegraphics[width=0.75\linewidth]{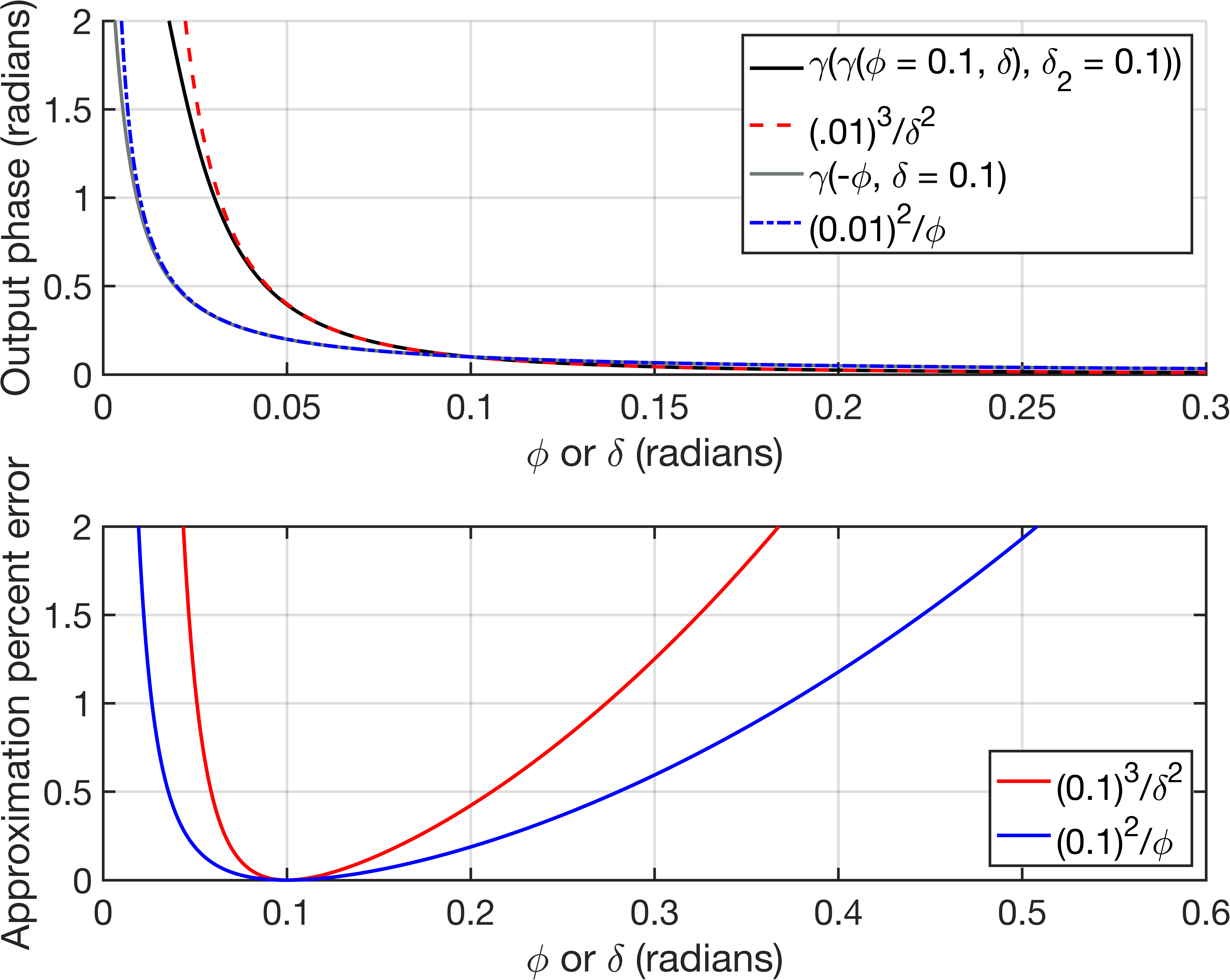}
    \caption{Illustration of a cascaded scalar operation. The parametrization $\gamma$ (Eq. (\ref{eq:gamma2})) is nested within itself, as shown in Fig. \ref{fig:rings}, which can produce the output phase $\gamma(\gamma(\phi, \delta_1), \delta_2)$. This can be biased to obtain an approximation to the nesting of the operation $\delta^2/\phi$ within itself, i.e. $\phi (\delta_2/\delta_1)^2$. Although cascading generally will result in error accumulation, this can be reduced by ensuring the inner operation's output falls close to the center of the output operation's most accurate region.} 
    \label{fig:cascade}
\end{figure}

Nevertheless, the present $\gamma$ is apt for cascading some operations without significant degradation in accuracy. For example, if the both the inner and outer parametrization is biased in the inversion regime, and $\phi_2$ is set to 0 modulo $2\pi$, the net output would be $\gamma(\gamma(\phi, \delta_1), \delta_2) \approx \gamma(\delta_1^2/(-\phi), \delta_2)$ for $\phi$ near $\delta_1$. The minus sign is acquired because it is $\gamma(-\phi, \delta)$ which approximates $\delta^2/\phi$; see Eqs. (\ref{eq:inv}). Next we assume $\delta_1^2/(-\phi)$ is near $\delta_2$, so that the outer division is valid to good approximation. This is in effect assuming $\phi$, $\delta_1$ and $\delta_2$ are all approximately the same. Applying the inversion approximation again, the end result is $\gamma(\delta_1^2/(-\phi), \delta_2) \approx \phi(\delta_2/\delta_1)^2$. This cascading has produced an approximation for a function proportional to the square of the quotient of two numbers. Fixing $\phi$ and $\delta_2$ allow approximations for $f(x) = 1/x^2$ to be obtained. An example of this is depicted in the top of Fig. \ref{fig:cascade} with corresponding error curves on the bottom. Overlaid with these for the sake of comparison is an example from the approximation of $\delta^2/\phi$.

Generally the nature of error accumulation from cascading depends on the individual functions involved and their own accuracies. Although a detailed study of this is beyond the present scope, we note here that in principle a variety of functions can be produced from repeated cascades of this $\gamma$. For example, nesting $\gamma$ in its second argument while fixing the value of the first argument repeatedly would allow the $f(x) = x^2$ approximation to be cascaded, so that an approximation for any even power in $x$ or $1/x$ is produced.

The increase in number of required phase shifter or beam-splitter resources resulting from cascading is found by counting the number of parameters introduced by the nesting process. For instance, if a one-argument function is nested continually within other one-argument functions, the total number of free parameters will stay constant. However, allowing for a translation of each output before it is passed to the next input will lead to a linear increase in phase parameters with the nesting depth. Similarily nesting two-argument functions within just one argument of each other will maintain a linear increase in required resources. However if multiple arguments of a parametrization are continually nested, the number of required resources will scale exponentially due to the exponential increase in free parameters. Hence, continually nesting two-argument functions within both arguments of each other will double the total number of arguments at each layer, leading to an exponential scaling of resources required. 

\section{Discussion and conclusion}
\begin{table}[htb!]
    \centering
    \begin{tabular}{c|c|c|c}
        Operation & Bias & Input & Output \\
        \hline
        Addition/subtraction & $\phi_0 = 0, \delta_0 = \pi/2$ & $\Delta\phi_1$, $\Delta\phi_2$ in series & $\Delta \gamma \approx \Delta\phi_1 \pm \Delta\phi_2$ \\
        Multiplication & $\phi_0 = 0$, $\delta_0 = 4.14$ (or nearby) & $\Delta\phi$, $\Delta\delta$ & $\Delta\gamma \approx \Delta\phi\Delta\delta$ \\
        Scalar inversion & $\phi_0 = \delta_0$ & $\Delta\phi$ & $\Delta\gamma \approx 1/(\phi_0 + \Delta\phi)$\\
        Squares & $\phi_0 = \delta_0$ & $\Delta\delta$ &  $\Delta\gamma \approx (\delta_0 + \Delta\delta)^2$
    \end{tabular}
    \caption{Summary of operations enacted by perturbing the phase function $\gamma(\phi, \delta)$ defined in Eq. (\ref{eq:gamma2}). All operation outputs are defined up to an overall scale factor, which is omitted in the last column.}
    \label{tab:table1}
\end{table}
The analog primitive operations introduced in this work are built on the use of small changes in phase about a bias point. All operations considered were encoded entirely in a phase change  $\Delta \gamma$. These operations are summarized in Table \ref{tab:table1}. The computed output $\Delta\gamma$ was converted to a probability $\Delta P$ in the most direct manner that a simple two-beam interference readout would permit, but other readouts remain to be explored. Improved readouts could be used to reduce the overhead error associated with the conversion between phase and probability for large input phase displacements. This ultimately pertains to controlling the terms in the Taylor series for $P$ with the goal being to make $P$ as linear in $\gamma$ as possible. With the basic Mach-Zehnder readout employed throughout this article, the readout parameter $\theta$ was selected to attain linearity through second order. But it cannot do better than this: if $P \coloneqq \sin^2((\gamma(\phi, \dots) - \theta)/2)$, the first few derivatives of $P$ are
\begin{subequations}
\begin{align}
    \frac{\partial P}{\partial \phi} \coloneqq P'(\gamma, \theta) &= \frac12 \gamma' \sin(\gamma - \theta) \\
    P'' &= \frac12 ( \gamma'' \sin(\gamma - \theta) - \gamma'^2 \cos(\gamma - \theta))\\
    P''' &= \frac12 (\gamma''' - \gamma'^3) \sin(\gamma - \theta) + \frac32 \gamma' \gamma '' \cos(\gamma - \theta). 
\end{align}
\end{subequations}
Selecting the value $\theta = \gamma_0 - \pi/2 - \pi m$ for any integer $m$ will set the first two derivatives to half of the corresponding derivative in $\gamma$. But at third order, this readout breaks down: the derivatives of $\gamma$ start to mix as the coefficients of the sine and cosine functions of $\theta$, becoming inseparable. An improved readout might be able to isolate the corresponding derivative of $\gamma$ through higher orders than the current readout.

In general, this approach allows higher computational accuracy to be obtained by further limiting the maximum change in the input phase. This also reduces the expected power consumed by each phase control channel, which is a critical aspect of scaling to very high dimensional systems \cite{hamerly2024informationtheoreticlimitprogrammablephotonics}. Generally speaking, by the power-series methodology used to construct the approximation, the working range of the computations will increase as the threshold of maximum error tolerated is increased. The exact nature of this relationship will depend on the parametrization as well as the bias point used. The relevant information can be extracted from a plot of error as in Figs. \ref{fig:addition}, \ref{fig:multiplication}, and \ref{fig:division03}. For a given maximum error threshold $\epsilon > 0$, the corresponding working range is found by seeking the \textit{first} input points $\phi_1$ and $\phi_2$ on either side of the bias point which generate an error value (in absolute value) that intersects the horizontal line at $y = \epsilon$.

\begin{figure}
    \centering
    \includegraphics[width=0.75\linewidth]{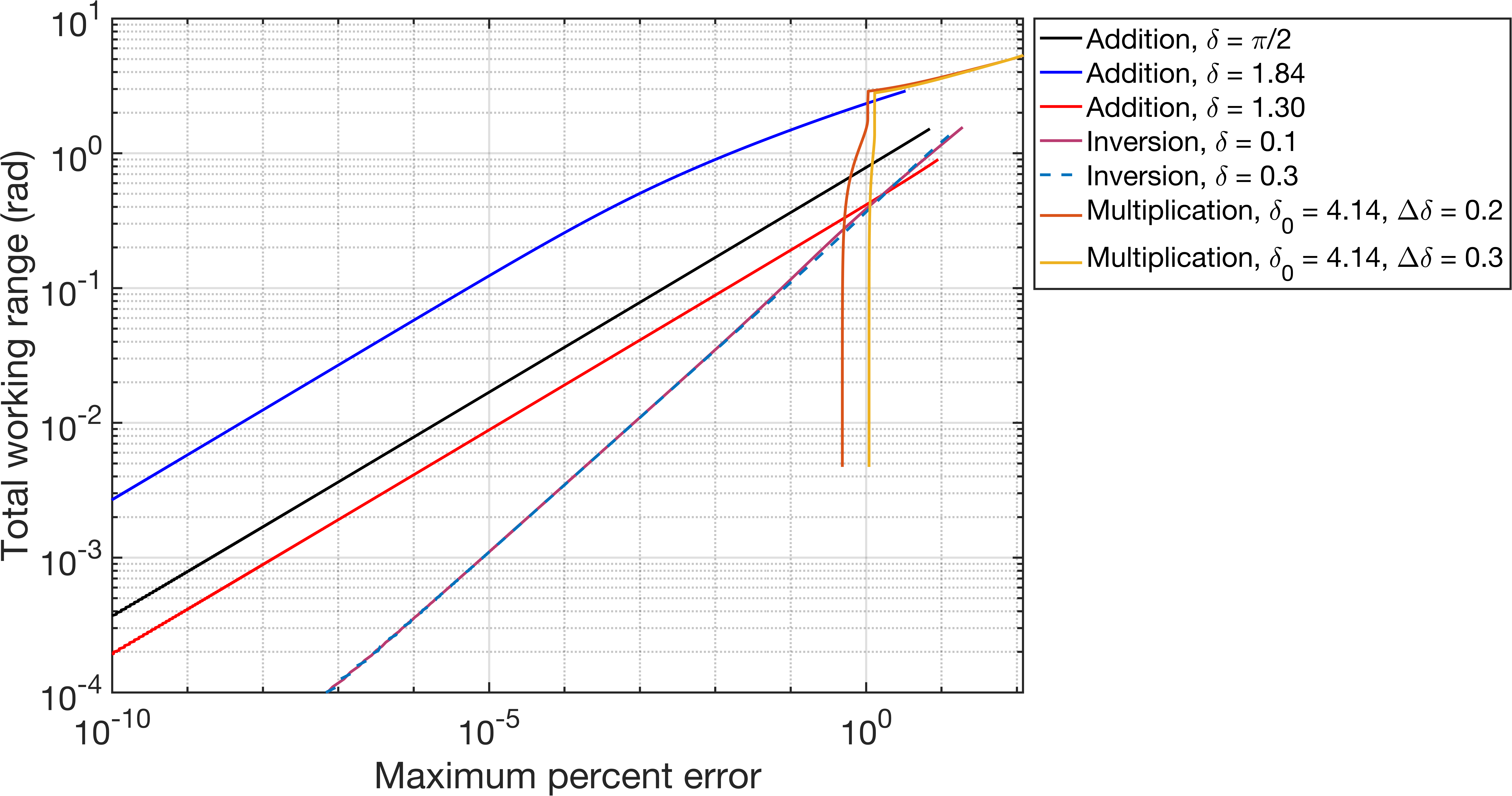}
    \caption{Dynamic range of input parameter $\phi$ vs. maximum percent error for the computations illustrated in prior sections, all originating from the same parametrization of Eq. (\ref{eq:gamma2}). The range for a given error is found by seeking the first parameter values on either side of the bias point which produce the error in question.}
    \label{fig:DR}
\end{figure}

This procedure was numerically applied to the error curves generated in the above plots to provide an illustration of some of the different working ranges this $\gamma$ can produce for its various operations. The corresponding maximum percent error vs. total working range curves are shown in Fig. \ref{fig:DR}. As expected, the range decreases with the error threshold. The inversion curves exhibit essentially the same range for different values of $\delta$ as described in the discussion surrounding Fig. \ref{fig:division03}. The multiplication curves, which use the central difference formulation, exhibit comparatively erroneous behavior, bottoming out on the order of 1\% error. This arises from the multiple approximations involved in the same computation, one in $\phi$ and another in $\delta$, each accumulating errors. Hence, convergence to zero error requires reducing the magnitude of $\Delta\delta$ as well as $\Delta \phi$. The basic calibration procedure discussed above can be used to ameliorate this. 

Encoding over multiple scattering states instead of a fixed state allowed the structure of the parameter space to influence the computation, allowing practical nonlinear operations to be obtained. These differential-phase encoded operations also have the benefit of being directly encoded by the physically applied parameter such as voltage or temperature. This circumvents the need for a complete reconfiguration algorithm to establish each desired operator, as in traditional unitary mesh schemes. In the present approach, only a one-time calibration is required to establish the correct bias points. But, if a system without active phase stabilization undergoes significant bias drift, e.g. from thermal fluctuations, it must be recalibrated. While the calibration holds, reconfiguration of the entire system can occur in a single pass, after adjusting each phase parameter once. In mesh schemes, rounds of iteration must occur to establish the desired operation. 

Unknown shifts in bias could generate major a detriment to accuracy. Noise in the phase controllers and/or optical source, if not properly averaged away, could add error on top of the approximation error. In conjunction with this, signficant deviation from the ideal system performance due to fabrication defects or loss could affect the accuracy of some operations to varying degrees. In some cases, the deviations may partially cancel out, due to the differential nature of the encoding. These practical aspects remain to be explored elsewhere.

A connection to finite differences was examined when constructing approximations for linear operations. This connection provides a direct approach for further generalizing the present results to control higher-order expansion terms when constructing a generic approximant $\Delta P$. This might be envisioned to obtain higher-order corrections to low-order terms, low-order term cancellations for the sake of obtaining higher-order nonlinear terms in isolation, and other general tailoring of more complicated approximations. However, this would invariably come at the expense of more terms in the definition of $\Delta P$, increasing the readout complexity and resource count.

Universal binary logic gates can be constructed from arithmetic operations. Addition translates to OR, multiplication to AND, and scalar inversion to NOT. So, despite the present technique being fully analog, it could be cast into a discrete formalism in order to combat noise and error accumulation. A variety of proposed optical computing schemes are based on a digital encoding \cite{10598302, Chen:24, Ashtiani:25, ZHANG2024130656, 8748932, 8360524, ghosh2024phasesymmetrybreakingcounterpropagating, Sharma2021}. In these approaches, logic levels are typically established by a shift in the resonant frequency of a tranmission dip or peak. In the present approach, however, all levels are in the value of the phase at a particular wavelength. This allows the use of multiple optical frequencies to be reserved for uses such as parallel processing channels, redundancy encoding, time-binning, etc.

In this work, we have introduced several different computationally-oriented operating regimes of a particular optical phase amplification map, Eq. (\ref{eq:gamma2}). The focus on a single map is to illustrate the basic operating principle and highlight its flexibility in producing different scalar operations. Certainly other interferometric configurations might be produced which have their own phase mapping characteristics, which in turn may produce a greater variety of higher accuracy computations within this paradigm. Later the approach will be generalized to matrix and vector operations.

The map of Eq. (\ref{eq:gamma2}) is realized with relatively common yet versatile components such as the beam-splitter and ring resonator. In the context of optical information processing, these devices have conventionally been crucial for modulation, filtering, and multiplexing systems. Common configurations using these have also been proposed for the sake of realizing various nonlinear activation functions for optical neural networks \cite{10128697, 9763349, 8769881}. Finally, we highlight that this is but one computing scheme that takes advantage of the potentially rich mapping between the parameter space and computation space of a linear interferometer. Other potentially nonperturbative computing schemes might be envisioned which harness this mapping as well. 

\section*{Funding.}
Air Force Office of Scientific Research MURI award number FA9550-22-1-0312.
\section*{Disclosures.}
None to report.
\section*{Data availability.}
All data generated for this article is available upon reasonable request to the authors.
\bibliography{refs}

\begin{thebibliography}{10}
\newcommand{\enquote}[1]{``#1''}

\bibitem{McMahon2023}
P.~L. McMahon, \enquote{The physics of optical computing,}
  {\protect\JournalTitle{Nature Reviews Physics}} \textbf{5}, 717--734 (2023).

\bibitem{StroevReview}
N.~Stroev and N.~G. Berloff, \enquote{Analog photonics computing for
  information processing, inference, and optimization,}
  {\protect\JournalTitle{Advanced Quantum Technologies}} \textbf{6}, 2300055
  (2023).

\bibitem{Athale:82}
R.~A. Athale and W.~C. Collins, \enquote{Optical matrix--matrix multiplier
  based on outer productdecomposition,} {\protect\JournalTitle{Appl. Opt.}}
  \textbf{21}, 2089--2090 (1982).

\bibitem{10.1063/5.0235712}
J.~R. Rausell-Campo, D.~Pérez-López, and J.~Capmany~Francoy,
  \enquote{Programming universal unitary transformations on a general-purpose
  silicon photonic platform,} {\protect\JournalTitle{APL Photonics}}
  \textbf{10}, 026102 (2025).

\bibitem{10.1117/12.2028585}
L.~Yang, L.~Zhang, and R.~Ji, \enquote{{On-chip optical matrix-vector
  multiplier},} in \emph{Optics and Photonics for Information Processing VII,}
  vol. 8855 K.~M. Iftekharuddin, A.~A.~S. Awwal, and A.~M{\'a}rquez, eds.,
  International Society for Optics and Photonics (SPIE, 2013), p. 88550F.

\bibitem{Gruber:00}
M.~Gruber, J.~Jahns, and S.~Sinzinger, \enquote{Planar-integrated optical
  vector-matrix multiplier,} {\protect\JournalTitle{Appl. Opt.}} \textbf{39},
  5367--5373 (2000).

\bibitem{Spall:20}
J.~Spall, X.~Guo, T.~D. Barrett, and A.~I. Lvovsky, \enquote{Fully
  reconfigurable coherent optical vector--matrix multiplication,}
  {\protect\JournalTitle{Opt. Lett.}} \textbf{45}, 5752--5755 (2020).

\bibitem{Tang:25}
R.~Tang, M.~Okano, C.~Zhang, K.~Toprasertpong, S.~Takagi, and M.~Takenaka,
  \enquote{Waveguide-multiplexed photonic matrix-vector multiplication
  processor using multiport photodetectors,} {\protect\JournalTitle{Optica}}
  \textbf{12}, 812--820 (2025).

\bibitem{Shen2017}
Y.~Shen, N.~C. Harris, S.~Skirlo, M.~Prabhu, T.~Baehr-Jones, M.~Hochberg,
  X.~Sun, S.~Zhao, H.~Larochelle, D.~Englund, and M.~Solja{\v{c}}i{\'{c}},
  \enquote{Deep learning with coherent nanophotonic circuits,}
  {\protect\JournalTitle{Nature Photonics}} \textbf{11}, 441--446 (2017).

\bibitem{Wang2022}
T.~Wang, S.-Y. Ma, L.~G. Wright, T.~Onodera, B.~C. Richard, and P.~L. McMahon,
  \enquote{An optical neural network using less than 1 photon per
  multiplication,} {\protect\JournalTitle{Nature Communications}} \textbf{13},
  123 (2022).

\bibitem{Filipovich:22}
M.~J. Filipovich, Z.~Guo, M.~Al-Qadasi, B.~A. Marquez, H.~D. Morison, V.~J.
  Sorger, P.~R. Prucnal, S.~Shekhar, and B.~J. Shastri, \enquote{Silicon
  photonic architecture for training deep neural networks with direct feedback
  alignment,} {\protect\JournalTitle{Optica}} \textbf{9}, 1323--1332 (2022).

\bibitem{Hua2025}
S.~Hua, E.~Divita, S.~Yu, B.~Peng, C.~Roques-Carmes, Z.~Su, Z.~Chen, Y.~Bai,
  J.~Zou, Y.~Zhu, Y.~Xu, C.-k. Lu, Y.~Di, H.~Chen, L.~Jiang, L.~Wang, L.~Ou,
  C.~Zhang, J.~Chen, W.~Zhang, H.~Zhu, W.~Kuang, L.~Wang, H.~Meng, M.~Steinman,
  and Y.~Shen, \enquote{An integrated large-scale photonic accelerator with
  ultralow latency,} {\protect\JournalTitle{Nature}} \textbf{640}, 361--367
  (2025).

\bibitem{doi:10.1126/sciadv.ads4224}
Q.~Feng, C.~B. Uzundal, R.~Guo, C.~Sanborn, R.~Qi, J.~Xie, J.~Zhang, J.~Wu, and
  F.~Wang, \enquote{Femtojoule optical nonlinearity for deep learning with
  incoherent illumination,} {\protect\JournalTitle{Science Advances}}
  \textbf{11}, eads4224 (2025).

\bibitem{Ahmed2025}
S.~R. Ahmed, R.~Baghdadi, M.~Bernadskiy, N.~Bowman, R.~Braid, J.~Carr, C.~Chen,
  P.~Ciccarella, M.~Cole, J.~Cooke, K.~Desai, C.~Dorta, J.~Elmhurst,
  B.~Gardiner, E.~Greenwald, S.~Gupta, P.~Husbands, B.~Jones, A.~Kopa, H.~J.
  Lee, A.~Madhavan, A.~Mendrela, N.~Moore, L.~Nair, A.~Om, S.~Patel, R.~Patro,
  R.~Pellowski, E.~Radhakrishnani, S.~Sane, N.~Sarkis, J.~Stadolnik,
  M.~Tymchenko, G.~Wang, K.~Winikka, A.~Wleklinski, J.~Zelman, R.~Ho, R.~Jain,
  A.~Basumallik, D.~Bunandar, and N.~C. Harris, \enquote{Universal photonic
  artificial intelligence acceleration,} {\protect\JournalTitle{Nature}}
  \textbf{640}, 368--374 (2025).

\bibitem{doi:10.1126/sciadv.adg7904}
L.~Bernstein, A.~Sludds, C.~Panuski, S.~Trajtenberg-Mills, R.~Hamerly, and
  D.~Englund, \enquote{Single-shot optical neural network,}
  {\protect\JournalTitle{Science Advances}} \textbf{9}, eadg7904 (2023).

\bibitem{PhysRevApplied.15.054034}
Y.~Zuo, Y.~Zhao, Y.-C. Chen, S.~Du, and J.~Liu, \enquote{Scalability of
  all-optical neural networks based on spatial light modulators,}
  {\protect\JournalTitle{Phys. Rev. Appl.}} \textbf{15}, 054034 (2021).

\bibitem{Bogaerts2020}
W.~Bogaerts, D.~P{\'e}rez, J.~Capmany, D.~A.~B. Miller, J.~Poon, D.~Englund,
  F.~Morichetti, and A.~Melloni, \enquote{Programmable photonic circuits,}
  {\protect\JournalTitle{Nature}} \textbf{586}, 207--216 (2020).

\bibitem{Nakajima2021}
M.~Nakajima, K.~Tanaka, and T.~Hashimoto, \enquote{Scalable reservoir computing
  on coherent linear photonic processor,} {\protect\JournalTitle{Communications
  Physics}} \textbf{4}, 20 (2021).

\bibitem{onodera2024scalingonchipphotonicneural}
T.~Onodera, M.~M. Stein, B.~A. Ash, M.~M. Sohoni, M.~Bosch, R.~Yanagimoto,
  M.~Jankowski, T.~P. McKenna, T.~Wang, G.~Shvets, M.~R. Shcherbakov, L.~G.
  Wright, and P.~L. McMahon, \enquote{Scaling on-chip photonic neural
  processors using arbitrarily programmable wave propagation,}  (2024).

\bibitem{PhysRevX.9.021032}
R.~Hamerly, L.~Bernstein, A.~Sludds, M.~Solja\ifmmode \check{c}\else
  \v{c}\fi{}i\ifmmode~\acute{c}\else \'{c}\fi{}, and D.~Englund,
  \enquote{Large-scale optical neural networks based on photoelectric
  multiplication,} {\protect\JournalTitle{Phys. Rev. X}} \textbf{9}, 021032
  (2019).

\bibitem{SchwarzeJB}
C.~R. Schwarze, D.~S. Simon, A.~D. Manni, A.~Ndao, and A.~V. Sergienko,
  \enquote{Finite-element assembly approach of optical quantum walk networks,}
  {\protect\JournalTitle{J. Opt. Soc. Am. B}} \textbf{41}, 1304--1316 (2024).

\bibitem{guggenheimer}
H.~Guggenheimer, \emph{Differential Geometry}, Dover Books on Advanced
  Mathematics (McGraw-Hill, 1963).

\bibitem{PhysRevA.107.052615}
C.~R. Schwarze, D.~S. Simon, and A.~V. Sergienko, \enquote{Enhanced-sensitivity
  interferometry with phase-sensitive unbiased multiports,}
  {\protect\JournalTitle{Phys. Rev. A}} \textbf{107}, 052615 (2023).

\bibitem{Schwarze:24}
C.~R. Schwarze, D.~S. Simon, A.~D. Manni, A.~Ndao, and A.~V. Sergienko,
  \enquote{Experimental demonstration of a grover-michelson interferometer,}
  {\protect\JournalTitle{Opt. Express}} \textbf{32}, 34116--34127 (2024).

\bibitem{PhysRevA.109.053508}
C.~R. Schwarze, D.~S. Simon, A.~Ndao, and A.~V. Sergienko, \enquote{Tunable
  linear-optical phase amplification,} {\protect\JournalTitle{Phys. Rev. A}}
  \textbf{109}, 053508 (2024).

\bibitem{Yang2024}
Y.~Yang, T.~Weiss, H.~Arianfard, A.~Youssry, and A.~Peruzzo, \enquote{A fixed
  phase tunable directional coupler based on coupling tuning,}
  {\protect\JournalTitle{Scientific Reports}} \textbf{14}, 24291 (2024).

\bibitem{Schwarze:25}
C.~R. Schwarze, A.~D. Manni, D.~S. Simon, A.~Ndao, and A.~V. Sergienko,
  \enquote{Grover-sagnac interferometer,} {\protect\JournalTitle{J. Opt. Soc.
  Am. A}} \textbf{42}, 290--297 (2025).

\bibitem{10.1063/1.5126517}
Z.~Ying, C.~Feng, Z.~Zhao, R.~Soref, D.~Pan, and R.~T. Chen,
  \enquote{Integrated multi-operand electro-optic logic gates for optical
  computing,} {\protect\JournalTitle{Applied Physics Letters}} \textbf{115},
  171104 (2019).

\bibitem{d140ec05d7604fad9e94354f7758a18e}
C.~Feng, J.~Gu, H.~Zhu, S.~Ning, R.~Tang, M.~Hlaing, J.~Midkiff, S.~Jain,
  D.~Pan, and R.~Chen, \enquote{Integrated multi-operand optical neurons for
  scalable and hardware-efficient deep learning,}
  {\protect\JournalTitle{Nanophotonics}}  (2024).

\bibitem{Yang2024-2}
G.~Yang, H.~Wang, S.~Mu, H.~Xie, T.~Wang, C.~He, M.~Shen, M.~Liu, C.~G. Van~de
  Walle, and H.~X. Tang, \enquote{Unveiling the pockels coefficient of
  ferroelectric nitride scaln,} {\protect\JournalTitle{Nature Communications}}
  \textbf{15}, 9538 (2024).

\bibitem{WIECHMANN20096847}
S.~Wiechmann and J.~Müller, \enquote{Thermo-optic properties of tio2, ta2o5
  and al2o3 thin films for integrated optics on silicon,}
  {\protect\JournalTitle{Thin Solid Films}} \textbf{517}, 6847--6849 (2009).

\bibitem{hamerly2024informationtheoreticlimitprogrammablephotonics}
R.~Hamerly, J.~R. Basani, A.~Sludds, S.~K. Vadlamani, and D.~Englund,
  \enquote{Towards the information-theoretic limit of programmable photonics,}
  (2024).

\bibitem{10598302}
S.~Ning, H.~Zhu, C.~Feng, J.~Gu, Z.~Jiang, Z.~Ying, J.~Midkiff, S.~Jain, M.~H.
  Hlaing, D.~Z. Pan, and R.~T. Chen, \enquote{Photonic-electronic integrated
  circuits for high-performance computing and ai accelerators,}
  {\protect\JournalTitle{Journal of Lightwave Technology}} \textbf{42},
  7834--7859 (2024).

\bibitem{Chen:24}
F.~Chen, S.~Zhou, Y.~Xia, X.~Yu, J.~Liu, F.~Li, and X.~Sui,
  \enquote{Ultra-compact optical full-adder based on directed logic and
  microring resonators,} {\protect\JournalTitle{Appl. Opt.}} \textbf{63},
  147--153 (2024).

\bibitem{Ashtiani:25}
F.~Ashtiani, \enquote{Programmable photonic latch memory,}
  {\protect\JournalTitle{Opt. Express}} \textbf{33}, 3501--3510 (2025).

\bibitem{ZHANG2024130656}
D.~Zhang, Z.~Fan, Y.~Dan, T.~Zhang, J.~Dai, and K.~Xu, \enquote{Bit-tunable xor
  operation based on photonic spiking neuron,} {\protect\JournalTitle{Optics
  Communications}} \textbf{565}, 130656 (2024).

\bibitem{8748932}
N.~Moroney, L.~Del~Bino, M.~T.~M. Woodley, J.~Silver, G.~N. Ghalanos, A.~Svela,
  S.~Zhang, and P.~Del'Haye, \enquote{Logic gates based on interaction of
  counterpropagating light in microresonators,} in \emph{2019 Conference on
  Lasers and Electro-Optics (CLEO),}  (2019), pp. 1--2.

\bibitem{8360524}
Z.~Ying, S.~Dhar, Z.~Zhao, C.~Feng, R.~Mital, C.-J. Chung, D.~Z. Pan, R.~A.
  Soref, and R.~T. Chen, \enquote{Electro-optic ripple-carry adder in
  integrated silicon photonics for optical computing,}
  {\protect\JournalTitle{IEEE Journal of Selected Topics in Quantum
  Electronics}} \textbf{24}, 1--10 (2018).

\bibitem{ghosh2024phasesymmetrybreakingcounterpropagating}
A.~Ghosh, A.~Pal, S.~Zhang, L.~Hill, T.~Bi, and P.~Del'Haye, \enquote{Phase
  symmetry breaking of counterpropagating light in microresonators for switches
  and logic gates,}  (2024).

\bibitem{Sharma2021}
S.~Sharma and S.~Roy, \enquote{Design of all-optical parallel multipliers using
  semiconductor optical amplifier-based mach--zehnder interferometers,}
  {\protect\JournalTitle{The Journal of Supercomputing}} \textbf{77},
  7315--7350 (2021).

\bibitem{10128697}
C.~Pappas, S.~Kovaios, M.~Moralis-Pegios, A.~Tsakyridis, G.~Giamougiannis,
  M.~Kirtas, J.~Van~Kerrebrouck, G.~Coudyzer, X.~Yin, N.~Passalis, A.~Tefas,
  and N.~Pleros, \enquote{Programmable tanh-, elu-, sigmoid-, and sin-based
  nonlinear activation functions for neuromorphic photonics,}
  {\protect\JournalTitle{IEEE Journal of Selected Topics in Quantum
  Electronics}} \textbf{29}, 1--10 (2023).

\bibitem{9763349}
J.~R. Rausell~Campo and D.~Pérez-López, \enquote{Reconfigurable activation
  functions in integrated optical neural networks,} {\protect\JournalTitle{IEEE
  Journal of Selected Topics in Quantum Electronics}} \textbf{28}, 1--13
  (2022).

\bibitem{8769881}
I.~A.~D. Williamson, T.~W. Hughes, M.~Minkov, B.~Bartlett, S.~Pai, and S.~Fan,
  \enquote{Reprogrammable electro-optic nonlinear activation functions for
  optical neural networks,} {\protect\JournalTitle{IEEE Journal of Selected
  Topics in Quantum Electronics}} \textbf{26}, 1--12 (2020).

\end{thebibliography}
\end{document}